# Purely electrical damping of vibrations in arbitrary PEM plates: a mixed non-conforming FEM-Runge-Kutta time evolution analysis

**F. dell'Isola, E. Santini, D. Vigilante**

**Summary** A new numerical code, based upon a mixed FEM-Runge-Kutta method, is used for the analysis and design of plane 2-D smart structures. The code is applied to the study of arbitrarily shaped PEM plates, based on a weak formulation of their governing equations, [17]. The optimal parameters needed to synthesize appropriate electric networks are computed, and the overall performances of such plates are investigated. Two examples are studied: firstly, a simple case is used to test the main features of the code; secondly, a more complex PEM plate is designed and analyzed by means of the proposed numerical approach.

**Keywords** Piezoelectricity, Smart structure, PEM plate, FEM, Optimization, Vibration damping, Runge-Kutta algorithm

## 1
## Introduction

A growing interest towards the possibilities opened by the concept of smart structures in industrial applications has arised recently, due to the innovations introduced by the material engineering in control techniques. There is the possibility now to design integrated systems, in which the control subsystem is no more based upon the paradigm of sensing-evaluating-actuating, but would be embedded in the system. Self-controlled structures, like piezo-electromechanical (PEM) beams and plates proposed in [14–17], seem to show very interesting performances, as it is suggested by theoretical analysis. The design of this class of smart structures is directed by the passive, purely electrical damping of mechanically excited structural vibrations. The main feature of the considered structures is that their evolution is governed by a system of PDEs where the unknown fields describe the electrical and mechanical state. The electrical state is not driven by an active control system and,

*Received 8 April 2002; accepted for publication 25 February 2003*

F. dell'Isola (✉)
Università degli studi di Roma "La Sapienza",
Dipartimento di ingegneria strutturale e geotecnica,
Via Eudossiana 18, 00184 Rome, Italy
e-mail: francesco.dell'isola@uniroma1.it

E. Santini
Università degli studi di Roma "La Sapienza",
Dipartimento di ingegneria elettrica,
Via Eudossiana 18, 00184 Rome, Italy
e-mail: ezio@elettrica.ing.uniroma1.it

D. Vigilante
Virginia Polytechnic Institute and State University,
Engineering Science and Mechanics Department,
24060 Blacksburg, Virginia, USA
e-mail: dvigilan@vt.edu

The ideas developed in this paper were discussed with Prof. Vincenzo Ciampi, Stefano Vidoli, PhD, Corrado Maurini, MS, and Maurizio Porfiri, MS. Their friendly help and constructive criticism were really useful.

therefore, no external "intelligence" is required in order to operate the mechanical vibration damping.

Obviously, the complexity of PEM structures does not allow for the determination of analytical or semi-analytical solutions of the governing PDEs in the general case. In order to supply a designing tool for experimental and industrial applications, a powerful enough numerical method for the analysis of such systems needs to be developed. There is an obvious difficulty in applying the FEM here, which has to be considered in this context: the finite elements to be introduced when modelling PEM plates must have at most a size comparable with the typical size of piezoelectric patches used. Indeed, a more refined mesh is clearly useless, as the homogenized equations which we derive are meaningful only for modes with wavelength greater than the upper bound of patch sizes.

In this paper, a mixed space-FEM and time-Runge-Kutta (RK) numerical code developed by the authors is described. It is applied to the analysis of the vibrational behavior of 2-D PEM plates. The mechanical part of these systems is described by the Kirchhoff–Love plate theory, while the homogenized electric system is governed by membrane-like equations.

As the electrical and mechanical states of considered PEM plates evolve, governed by a second- and fourth-order in space-differential operators, respectively, the main difficulty in the code development consisted in the determination of a set of shape-functions suitable to represent both the aforementioned states. This difficulty has been confronted following the ideas described in [1] for the generic fourth-order PDE: the main trick is to build and use finite elements starting from a *non-conforming* triangular element with three corner nodes. The resulting FE approximation has the following properties:

i) as it passes the *patch test*, it converges monotonically in the energy norm (the convergence which needs to be proven case-by-case for non-conforming elements);
ii) as it includes only three corner nodes, it is immediately adapted to represent approximate solutions of second-order space PDEs.

The novelty of the proposed FEM code consists in the simultaneous solution of the studied coupled PDEs by means of the same mesh of elements.

The subsequent time-integration presents great numerical difficulties: indeed, using the FE projection, the governing system of PDEs is reduced to a system of time ODEs, the dimension of which can be huge. Therefore, in order to make the calculation possible or to save some computational time, attention must be paid to the implementation of the RK algorithm: in order to reduce the dimension of the time ODEs system, its modal representation has been used and truncated at a proper number of electric and mechanical modes.

The numerical code is then used to study the electromechanical behavior of a specific PEM plate with complex geometry. In particular, the design of such plate is performed by means of an optimal-parameters evaluation procedure, using the developed code as a subroutine.

More precisely, once the mechanical properties of the considered PEM plates are fixed, the proposed code allows for the determination of net inductances and net resistances, which optimizes electrical damping of the vibrational energy of a specific vibrational mode. This is done by tuning the electrical network to the desired mechanical frequency, i.e. choosing the net inductance, and subsequently minimizing the mechanical damping ratio, i.e. choosing the net resistance. We explicitly remark that, in our model, the inertia and stiffness of the introduced piezoelectric actuators can be accounted for by suitably modifying the homogenized constitutive parameters of the PEM plate. In the numerical symulations shown in the present paper, the influence of the added stiffness of piezoelectric actuators has been considered when evaluating the PEM constitutive parameters, while the added mass has been assumed as (and for the considered plate and actuators effectively is) negligible.

As generally proven in [16], every structural member type of vibration (e.g. flexural, torsional or in-plane) needs to be coupled via suitable piezoelectric actuators to a suitably tuned electric circuit. In the present paper, we address the problem of controlling flexural vibrations which, in view of the results shown in [16], is more difficult than the problem of controlling in-plane vibrations. Indeed, in-plane vibrations are governed by second-order PDEs and can be easily coupled to a standard transmission network (for more details see [17]).

It is finally shown that the specific designed device is able to damp mechanically induced vibrations in a specific range of frequencies.

## 2
## Description of the system

The problem of controlling vibrations in beams and plates by distributing on them a set of piezoelectric (PZT) actuators (suitably electrically interconnected) has been addressed in [14–17]. A new concept of 2-D electrical continuum was introduced, the evolution of which parallels that one valid for the plate.

The main goal of this concept is to synthesize an electric circuit (which is coupled with the mechanical structure through piezo-actuators) and which would behave analogous to the mechanical system. Exploiting this concept, it has been proposed to control plate-like structures by means of an internal resonance phenomenon coupling mechanical and electrical modes.

Actually, we use discrete components, which connect a finite number of piezoelectric actuators which become Two Terminal Networks (TTN) belonging to a modular electric network. According to the hypothesis of large wavelengths as compared with the dimensions of components and modules (this is a plausible hypothesis, considering the mechanical wavelengths of lower modes), it is possible to study the electric network by means of a continuum (homogeneized) model. Therefore, we consider a 2-D generalization of the transmission line. Using the well-known synthesis techniques of discrete electric network, it is possible to simulate every differential equation (for a detailed discussion of this point see [20]).

In practice, in order to obtain the maximum coupling on every vibration mode, we want the dispersive relations that relate group velocity and frequency to be the same in the electric and mechanical subsystems, see [16]. To satisfy this requirement, the electric networks should *exhibit* the same governing differential equations that rule the mechanical subsystem. We choose the electric components in order to slow down the group velocity of the electric wave so as to make it equal to the mechanical one. In this way, every electromechanical signal which propagates through the structure has a substantial electric content, which can easily be managed by the control system.

We remark explicitly that we aim to dissipate in a purely electrical way the mechanical vibrational energy. This can be done by means of electrical resistors that in the chosen topology (Fig. 1) can be in series to the inductances or in parallel to the PZT.

Obviously, the electric network must be adapted depending on the mechanical phenomena we wish to control: for example, the longitudinal and torsional vibrations of a bar, governed by second-order differential equations, or the flexural vibrations, governed by fourth-order differential equations, are to be damped with dedicated electric networks.

Only some elementary structures (as plates or beams with simple shapes) have been investigated untill now using semi-analytical methods and developing a suitable electric network for interconnecting the actuators. In particular, for flexural waves, the synthesis of an analogous circuit governed by fourth-order equations is not straightforward, s. [18], [20] and [19].

Nevertheless, some very interesting results can be obtained by controlling the structure with a second-order electric network. In this way, we can maximize the coupling of only one vibration mode, but the circuit we have to realize is very simple. Anyway, this control technique has proven to have better performances and lower costs than other PZT-based control techniques, [21, 22].

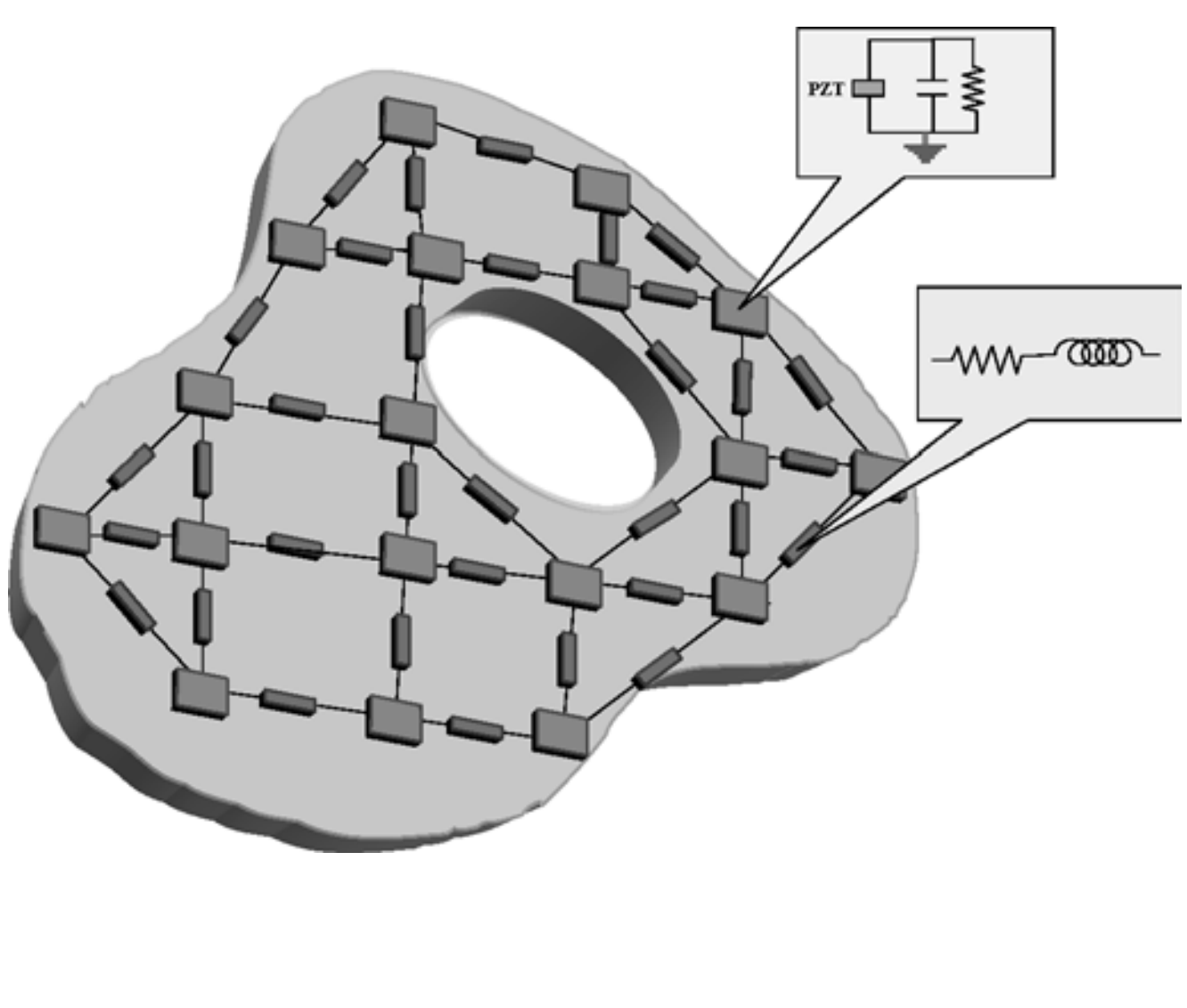

**Fig. 1.** Lumped approximation of an electromechanical continuum

The PEM plates obtained by interconnecting the electric terminals of the PZT transducers with a second-order electric network are governed by the following system of PDEs, [17]:

$$\begin{aligned} &\triangle\triangle w + \alpha \ddot{w} - \gamma \dot{\psi} = 0, \\ &\triangle\varphi - \beta\ddot{\varphi} + \gamma\triangle\dot{w} = 0 \ , \end{aligned} \tag{1}$$

to be completed by suitable boundary and initial conditions. Here, $w$ is the mechanical displacement, $\psi$ is a time-integral of the electric tension and $\gamma$ is the coupling factor; $\triangle$ holds for 2-D laplacian operator, while $\dot{x}$ denotes the time derivative of $x$. 

## 3 Weak formulation

In this section, we discuss the mathematical model for the systems under investigation. A weak formulation of the Eqs. (1) is derived, and the parameters needed by the numerical code are computed by means of a standard identification procedure. We determine the constitutive relations for PEM plates at our macrolevel in terms of the mechanical and electrical parameters characterizing the constituting mechanical and electrical elements at a microlevel. The identification in expended power simply generalizes the standard theory of Kirchhoff–Love plate, comp. [17].

We remark that, in order to describe the main features of the behavior of the investigated electromechanical system, a continuum model was introduced in [17]. Since the physical realization of the electric part of PEM plates is obtained by lumped elements, this model allows for an accurate description of vibrations only when the involved wavelengths are not too small compared with the dimensions of the single element and the circuital module. By formulating the balance of power for the considered PEM plate, we obtain a weak formulation of the problem (1), which allows for a numerical analysis by means of the FEM.

We deal with a plate body $\mathcal{B}$ occupying the region $\mathcal{C} = \mathcal{S} \times \mathcal{I}$, where $\mathcal{S}$ is a plane surface and $\mathcal{I}$ the real interval $[-h, h]$. As usual, the thickness $2h$ is supposed to be small compared to the diameter of $\mathcal{S}$. The electric network is assumed to be 2-D. The following balance of power:

$$\langle \mathbf{b}, \dot{\mathbf{u}}_t \rangle_{\mathcal{C}} + \langle \mathbf{f}, \dot{\mathbf{u}}_t \rangle_{\partial\mathcal{C}} = \langle \boldsymbol{\sigma}, \dot{\boldsymbol{\varepsilon}}_t \rangle_{\mathcal{C}} \ , \tag{2}$$

must hold for every compatible test field $(\dot{u}, \dot{\varepsilon})$. Here, the notation stands for

$$\langle \mathbf{x}, \mathbf{y} \rangle_{\mathcal{S}} = \int_{\mathcal{S}} \mathbf{x}^T \mathbf{y} \, d\mathcal{S} \ ,$$

$\mathbf{u}$ is the vector of kinematical descriptors of the system, $\boldsymbol{\varepsilon}$ the deformation field, $\mathbf{b}, \mathbf{B}_1, \mathbf{B}_2, \mathbf{s}_1, \mathbf{s}_2$ and $\mathbf{f}$ are the generalized body resp. surface external or inertial forces, $\boldsymbol{\sigma}$ is the internal stress field. The index $t$ indicates that we deal with a test field. We have

$$\mathbf{b} = \begin{bmatrix} b_1 \\ b_2 \\ b_z \\ i \end{bmatrix}, \quad \mathbf{u} = \begin{bmatrix} u \\ v \\ w \\ \psi \end{bmatrix}, \quad \mathbf{f} = \begin{bmatrix} f_1 \\ f_2 \\ f_z \\ \imath \end{bmatrix} \ ,$$

where $(b_1, b_2, b_z)$ and $(f_1, f_2, f_z)$ are the components of the mechanical external forces, $i$ and $\imath$ are body and surface current densities from the ground, $(u, v, w)$ are the components of the mechanical displacements vector and $\psi$ is the time-integral of the electric tension between a point on the surface and the reference terminal.

According to the Cauchy model and using Voigt representation we have the following stress and deformation fields

$$\boldsymbol{\sigma} = \begin{bmatrix} \sigma_1 \\ \sigma_2 \\ \sigma_z \\ \tau_{12} \\ \tau_{2z} \\ \tau_{z1} \\ I_1 \\ I_2 \end{bmatrix}, \qquad \boldsymbol{\varepsilon} = \begin{bmatrix} \varepsilon_1 \\ \varepsilon_2 \\ \varepsilon_z \\ \gamma_{12} \\ \gamma_{2z} \\ \gamma_{z1} \\ \Delta_1 \\ \Delta_2 \end{bmatrix} = \begin{bmatrix} u_{,1} \\ v_{,2} \\ w_{,z} \\ u_{,2} + v_{,1} \\ w_{,1} + u_{,z} \\ v_{,z} + w_{,2} \\ \psi_{,1} \\ \psi_{,2} \end{bmatrix},$$

where $(I_1, I_2)$ are the components of the surface current density and $()_{,i}$ stands for the space derivative in the $i$th direction within the plane.

According to the geometry of the body, the position vector in the reference configuration is decomposed as

$$\mathrm{x} = \mathbf{r} + z\mathbf{e} \ , \tag{3}$$

where $\mathbf{r}$ is the position vector in S, $z \in \mathcal{I}$ and $\mathbf{e}$ is the unit vector perpendicular to S. To deduce from the 3-D Cauchy model of B the behavior of a bending plate, we use the Kirchhoff–Love-compatibile identification procedure based on the following kinematical reduction map for mechanical displacements:

$$\mathbf{u}_m(\mathbf{r},\zeta) = \begin{bmatrix} -\frac{\partial w(\mathbf{r})}{\partial r_1} z \\ -\frac{\partial w(\mathbf{r})}{\partial r_2} z \\ w(\mathbf{r}) \end{bmatrix} .$$

Hence, the function $w(\mathbf{r})$ models the transverse displacements of the points of the plate. As it has been previously discussed, the electric system is assumed to be 2-D, and the voltage $\dot{\psi}$ depends on $\mathbf{r}$ only. Therefore

$$\mathbf{u}(\mathbf{r}) = \begin{bmatrix} \mathbf{u}_m(\mathbf{r}) \\ \psi(\mathbf{r}) \end{bmatrix} = \begin{bmatrix} -\frac{\partial w(\mathbf{r})}{\partial r_1} z \\ -\frac{\partial w(\mathbf{r})}{\partial r_2} z \\ w(\mathbf{r}) \\ \psi(\mathbf{r}) \end{bmatrix} . \tag{4}$$

Substituting in Eq. (2) the reduction map (4), we obtain

$$\left\langle \hat{\mathbf{b}}, \hat{\mathbf{u}} \right\rangle_{\mathcal{S}} + \langle \mathbf{B}_1, \hat{\mathbf{u}}_{,1} \rangle_{\mathcal{S}} + \langle \mathbf{B}_2, \hat{\mathbf{u}}_{,2} \rangle + \left\langle \hat{\mathbf{f}}, \hat{\mathbf{u}} \right\rangle_{\partial\mathcal{S}} + \langle \mathbf{s}_1, \hat{\mathbf{u}}_{,1} \rangle_{\partial\mathcal{S}} + \langle \mathbf{s}_2, \hat{\mathbf{u}}_{,2} \rangle_{\partial\mathcal{S}} = \langle \hat{\boldsymbol{\sigma}}, \hat{\boldsymbol{\varepsilon}} \rangle_{\mathcal{S}} \ , \tag{5}$$

where

$$\hat{\mathbf{u}} = \begin{bmatrix} w \\ \psi \end{bmatrix} \quad \hat{\mathbf{b}} = \begin{bmatrix} g \\ i \end{bmatrix}, \quad \hat{\mathbf{f}} = \begin{bmatrix} q \\ \imath \end{bmatrix},$$

$$\mathbf{B}_1 = \begin{bmatrix} B_1 \\ 0 \end{bmatrix}, \quad \mathbf{B}_2 = \begin{bmatrix} B_2 \\ 0 \end{bmatrix}, \quad \mathbf{s}_1 = \begin{bmatrix} s_1 \\ 0 \end{bmatrix},$$

$$\mathbf{s}_2 = \begin{bmatrix} s_2 \\ 0 \end{bmatrix}, \quad \hat{\boldsymbol{\sigma}} = \begin{bmatrix} m_1 \\ m_2 \\ m_{12} \\ I_1 \\ I_2 \end{bmatrix}, \quad \hat{\boldsymbol{\varepsilon}} = \begin{bmatrix} \chi_1 \\ \chi_2 \\ \chi_{12} \\ \psi_{,1} \\ \psi_{,2} \end{bmatrix},$$

are the dynamical action and the kinematical fields in the reduced plate model, and where the following definitions have been introduced:

$$
\begin{aligned}
& g(\mathbf{r}) = \int_{-h}^{h} b_z(\mathbf{r})\mathrm{d}z, \quad q(\mathbf{r}) = \int_{-h}^{h} f_z(\mathbf{r})\mathrm{d}z, \\
& B_1(\mathbf{r}) = \int_{-h}^{h} b_1(\mathbf{r})z\ \mathrm{d}z, \quad B_2(\mathbf{r}) = \int_{-h}^{h} b_2(\mathbf{r})z\ \mathrm{d}z, \\
& s_1(\mathbf{r}) = \int_{-h}^{h} f_1(\mathbf{r})z\ \mathrm{d}z, \quad s_2(\mathbf{r}) = \int_{-h}^{h} f_2(\mathbf{r})z\ \mathrm{d}z, \\
& m_1(\mathbf{r}) = \int_{-h}^{h} \sigma_1(\mathbf{r})z\ \mathrm{d}z, \quad m_2(\mathbf{r}) = \int_{-h}^{h} \sigma_2(\mathbf{r})z\ \mathrm{d}z, \\
& m_{12}(\mathbf{r}) = \int_{-h}^{h} \tau_{12}(\mathbf{r})z\ \mathrm{d}z, \\
& \chi_1 = -w_{,11}, \quad \chi_2 = -w_{,22}, \quad \chi_{12} = -2w_{,12} \ .
\end{aligned}
\tag{6}
$$

We can express the infinitesimal deformation field in the following way:

$$
\hat{\boldsymbol{\varepsilon}} = \begin{bmatrix} \chi_1 \\ \chi_2 \\ \chi_{12} \\ \psi_{,1} \\ \psi_{,2} \end{bmatrix} = \mathbf{H}_0\hat{\mathbf{u}} + \mathbf{H}_1\hat{\mathbf{u}}_{,1} + \mathbf{H}_2\hat{\mathbf{u}}_{,2} + \mathbf{H}_3\hat{\mathbf{u}}_{,11} + \mathbf{H}_4\hat{\mathbf{u}}_{,22} + \mathbf{H}_5\hat{\mathbf{u}}_{,12}
$$

To complete the model, we need to consider the constitutive relations. For the mechanical part, assuming that the body $\mathcal{B}$ is linear and orthotropic, we get

$$
\begin{bmatrix} \boldsymbol{\sigma}_1 \\ \boldsymbol{\sigma}_2 \end{bmatrix} = \begin{bmatrix} \mathbf{E}_1 & \mathbf{0} \\ \mathbf{0} & \mathbf{E}_2 \end{bmatrix} \begin{bmatrix} \boldsymbol{\varepsilon}_1 \\ \boldsymbol{\varepsilon}_2 \end{bmatrix} \ ,
$$

where $\boldsymbol{\sigma}_1^T = \{\sigma_1, \sigma_2, \sigma_z\}$ and $\boldsymbol{\varepsilon}_1^T = \{\varepsilon_1, \varepsilon_2, \varepsilon_z\}$. In the reduced model, we have

$$
\mathbf{m} = \mathbf{E}_\chi \chi \ ,
$$

where

$$
\mathbf{E}_\chi = \frac{2h^3}{3}\mathbf{E}_1 \ .
$$

Moreover, considering inertia forces, from (6) we get

$$
g = -2h\rho\ddot{w} + g_0 \ ,
$$

$$
B_1 = -\frac{2h^3}{3}\rho\ddot{w}_{,1} + B_1^0, \quad B_2 = -\frac{2h^3}{3}\rho\ddot{w}_{,2} + B_2^0 \ .
$$

Assuming the network to be linear and dissipative, the purely electric constitutive relations are

$$
\begin{aligned}
& i_e = K_c\ddot{\psi} + G_N\dot{\psi} + i_0, \\
& -\dot{\psi}_{,1} = L_N\frac{\partial I_1}{\partial t} + R_N I_1, \\
& -\dot{\psi}_{,2} = L_N\frac{\partial I_2}{\partial t} + R_N I_2 \ ,
\end{aligned}
$$

where $K_c$ and $G_N$ are, respectively, the capacitance and conductance from the surface to the ground, and $L_N$ and $R_N$ are, respectively, the net-inductance and net-resistance. We can invert the last two relations, getting

$$\mathbf{I} = \mathbf{C}_0 e^{-\frac{R_N}{L_N}t} - \frac{1}{L_N}\int_{t_0}^{t} e^{-\frac{R_N}{L_N}(t-\tau)}(\text{grad } \dot{\psi})\mathrm{d}\tau \ .$$

Now let's assume the current on the net to be zero for $t = t_0$, so that $\mathbf{C}_0 = \emptyset$. We can now define the following integral operator:

$$Y_L(.) = -\frac{1}{L_N}\int_{t_0}^{t} e^{-\frac{R_N}{L_N}(t-\tau)}(.)\mathrm{d}\tau \ ,$$

and assuming that time and space operators commute note that

$$\mathbf{I} = Y_L(\text{grad } \dot{\psi}) = \text{grad } Y_L(\dot{\psi}) \ .$$

In addition, we specify the constitutive relation of a single PZT actuator

$$\begin{bmatrix} m_1 \\ m_2 \\ m_{12} \\ \frac{Q}{d^2} \end{bmatrix} = \begin{bmatrix} g_{mm1} & 0 & 0 & -g_{me1} \\ 0 & g_{mm2} & 0 & -g_{me2} \\ 0 & 0 & g_{mm12} & -g_{me12} \\ g_{me1} & g_{me2} & g_{me12} & g_{ee} \end{bmatrix} \begin{bmatrix} \chi_1 \\ \chi_2 \\ \chi_{12} \\ \psi \end{bmatrix} ,$$

which can be rearranged as follows:

$$\begin{bmatrix} \mathbf{m} \\ \frac{Q}{d^2} \end{bmatrix} = \begin{bmatrix} \mathbf{E}_\chi^p & -\mathbf{E}_c^T \\ \mathbf{E}_c & g_{ee} \end{bmatrix} \begin{bmatrix} \boldsymbol{\chi} \\ \psi \end{bmatrix}$$

where $\mathbf{m}$ and $\chi$ are the bending moments and curvatures, while $Q/d^2$ and $\dot{\psi}$ are the charge per unit area and voltage between the actuators plates.

Therefore, the coupled constitutive relations for $\mathbf{m}$ and $i$ read as follows:

$$\begin{aligned} \mathbf{m} &= \left(\mathbf{E}_\chi + \mathbf{E}_\chi^p\right)\boldsymbol{\chi} - \mathbf{E}_c^T\dot{\psi}, \\ i &= i_e + \frac{\dot{Q}}{d^2} = (K_c + g_{ee})\ddot{\psi} + G_N\dot{\psi} + \mathbf{E}_c\dot{\boldsymbol{\chi}} + i_0 = C_N\ddot{\psi} + G_N\dot{\psi} + \mathbf{E}_c\dot{\boldsymbol{\chi}} + i_0 \ . \end{aligned}$$

We have now to face the problem of the time integral operator $Y_L$. In order to remove it, we have to change the electric kinematical descriptor in the following way. Let's define

$$Y_L\left(\dot{\psi}\right) := \alpha \ ,$$

hence

$$\begin{aligned} L_N\dot{\alpha} + R_N\alpha &= -\dot{\psi}, \\ L_N\ddot{\alpha} + R_N\dot{\alpha} &= -\ddot{\psi} \ . \end{aligned} \tag{7}$$

Therefore the overall constitutive relations read as follows:

$$\mathbf{m} = \mathbf{E}_{\chi g}\boldsymbol{\chi} + \mathbf{E}_c^T(L_N\dot{\alpha} + R_N\alpha),$$
$$i = -C_N(L_N\ddot{\alpha} + R_N\dot{\alpha}) - G_N(L_N\dot{\alpha} + R_N\alpha) + \mathbf{E}_c\dot{\boldsymbol{\chi}},$$
$$\mathbf{I} = \text{grad}\ \boldsymbol{\alpha}, \quad g = -2h\rho\ddot{w} + g_0,$$
$$B_1 = -\frac{2h^3}{3}\rho\ddot{w}_{,1} + B_1^0, \quad B_2 = -\frac{2h^3}{3}\rho\ddot{w}_{,2} + B_2^0,$$
$$\mathbf{s} = \mathbf{s}_0, \quad \hat{\mathbf{f}} = \hat{\mathbf{f}}_0 \ .$$

Finally we can provide the complete weak formulation in the following resume:

**Problem Balance of power**

$$\langle \mathbf{b}, \dot{\mathbf{u}}^t \rangle_{\mathcal{S}} + \left\langle \mathbf{B}_1, \dot{\mathbf{u}}^t_{,1} \right\rangle_{\mathcal{S}} + \left\langle \mathbf{B}_2, \dot{\mathbf{u}}^t_{,2} \right\rangle + \langle \mathbf{f}, \dot{\mathbf{u}}^t \rangle_{\partial\mathcal{S}} + \left\langle \mathbf{s}_1, \dot{\mathbf{u}}^t_{,1} \right\rangle_{\partial\mathcal{S}} + \left\langle \mathbf{s}_2, \dot{\mathbf{u}}^t_{,2} \right\rangle_{\partial\mathcal{S}} = \langle \boldsymbol{\sigma}, \dot{\boldsymbol{\varepsilon}}^t \rangle_{\mathcal{S}} \ ,$$

must hold for any compatible test field $(\mathbf{u}^t, \varepsilon^t)$:

**Constitutive relations and external-inertia forces**

$$\mathbf{b} = \mathbf{G}\ddot{\mathbf{u}} + \mathbf{S}\dot{\mathbf{u}} + \mathbf{T}\mathbf{u} + \mathbf{V}\dot{\boldsymbol{\varepsilon}} + \mathbf{b}_0,$$
$$\mathbf{B}_1 = \mathbf{G}_B^1\ddot{\mathbf{u}}_{,1} + \mathbf{B}_1^0, \quad \mathbf{B}_2 = \mathbf{G}_B^2\ddot{\mathbf{u}}_{,2} + \mathbf{B}_2^0,$$
$$\mathbf{f} = \mathbf{f}_0,$$
$$\mathbf{s}_1 = \mathbf{s}_1^0, \quad \mathbf{s}_2 = \mathbf{s}_2^0,$$
$$\boldsymbol{\sigma} = \mathbf{E}\boldsymbol{\varepsilon} + \mathbf{C}\dot{\mathbf{u}} + \mathbf{R}\mathbf{u} + \boldsymbol{\sigma}_0 \ .$$

**Kinematical compatibility**

$$\boldsymbol{\varepsilon} = \mathbf{H}_0\mathbf{u} + \mathbf{H}_1\mathbf{u}_{,1} + \mathbf{H}_2\mathbf{u}_{,2} + \mathbf{H}_3\mathbf{u}_{,11} + \mathbf{H}_4\mathbf{u}_{,22} + \mathbf{H}_5\mathbf{u}_{,12},$$
$$\boldsymbol{\varepsilon}^t = \mathbf{H}_0\mathbf{u}^t + \mathbf{H}_1\mathbf{u}^t_{,1} + \mathbf{H}_2\mathbf{u}^t_{,2} + \mathbf{H}_3\mathbf{u}^t_{,11} + \mathbf{H}_4\mathbf{u}^t_{,22} + \mathbf{H}_5\mathbf{u}^t_{,12} \ ,$$

where

$$\mathbf{G} = \begin{bmatrix} -2h\rho & 0 \\ 0 & -C_N L_N \end{bmatrix}, \quad \mathbf{S} = \begin{bmatrix} 0 & 0 \\ 0 & -C_N R_N - G_N L_N \end{bmatrix},$$
$$\mathbf{G}_B^1 = \begin{bmatrix} -\frac{2h^3}{3}\rho & 0 \\ 0 & 0 \end{bmatrix}, \quad \mathbf{G}_B^2 = \begin{bmatrix} -\frac{2h^3}{3}\rho & 0 \\ 0 & 0 \end{bmatrix},$$
$$\mathbf{T} = \begin{bmatrix} 0 & 0 \\ 0 & -G_N R_N \end{bmatrix}, \quad \mathbf{V} = \begin{bmatrix} g_{me1} & g_{me2} & g_{me12} & 0 & 0 \\ 0 & 0 & 0 & 0 & 0 \end{bmatrix},$$
$$\mathbf{E} = \begin{bmatrix} \mathbf{E}_{\chi g} & \mathbf{0} \\ \mathbf{0} & \mathbf{1} \end{bmatrix}, \quad \mathbf{R} = \begin{bmatrix} 0 & R_N\ g_{me1} \\ 0 & R_N\ g_{me2} \\ 0 & R_N\ g_{me12} \\ 0 & 0 \\ 0 & 0 \end{bmatrix}$$
$$\mathbf{C} = \begin{bmatrix} 0 & L_N\ g_{me1} \\ 0 & L_N\ g_{me2} \\ 0 & L_N\ g_{me12} \\ 0 & 0 \\ 0 & 0 \end{bmatrix} .$$

and

$$\mathbf{H}_0 = \begin{bmatrix} 0 & 0 \\ 0 & 0 \\ 0 & 0 \\ 0 & 0 \\ 0 & 0 \end{bmatrix}, \quad \mathbf{H}_1 = \begin{bmatrix} 0 & 0 \\ 0 & 0 \\ 0 & 0 \\ 0 & 1 \\ 0 & 0 \end{bmatrix},$$

$$\mathbf{H}_2 = \begin{bmatrix} 0 & 0 \\ 0 & 0 \\ 0 & 0 \\ 0 & 0 \\ 0 & 1 \end{bmatrix}, \quad \mathbf{H}_3 = \begin{bmatrix} -1 & 0 \\ 0 & 0 \\ 0 & 0 \\ 0 & 0 \\ 0 & 0 \end{bmatrix},$$

$$\mathbf{H}_4 = \begin{bmatrix} 0 & 0 \\ -1 & 0 \\ 0 & 0 \\ 0 & 0 \\ 0 & 0 \end{bmatrix}, \quad \mathbf{H}_5 = \begin{bmatrix} 0 & 0 \\ 0 & 0 \\ -2 & 0 \\ 0 & 0 \\ 0 & 0 \end{bmatrix} .$$

Let us observe that with the model proposed in the Problem, completed by adequate boundary and initial conditions, we can describe with a weak formulation 2-D coupled electrical and mechanical subsystems governed by second- and fourth-order differential equations, respectively, and that we can analyze any number of them coupled in different ways.

The data needed by the program are simply all the matrices needed in this representation, which, for the application under consideration, are those listed in this section.

Let us remark that the numerical procedure can be divided into two main parts: the space analysis, performed by means of a nonconforming FEM calculation (Galerkin approach), and the time-evolution analysis, numerically solved with a RK algorithm. For more details on this part see Appendix.

## 4 Numerical results

### 4.1 Test case: simply supported square PEM plate

We start considering a simply supported square PEM plate in order to obtain a first test of the performances of the developed code.

The case under consideration has been deeply investigated in [17], where semi-analitycal solutions have been determined by means of modal analysis. The advantage of this problem is that not only the uncoupled electric and mechanical systems are solved easily because of the simple geometry, but also for the coupled system, formed by a second-order system (the electric network) and a fourth-order one (the flexural mechanical vibrations), a simple series representation of the solution can be provided. We will show how the numerical process converges to the exact solution for both cases and how the coupled system is approximated as well.

#### 4.1.1 Uncoupled systems

The classical solutions for the simply supported square plate, as well as for the second-order (membrane-like) equations of the electric network, are compared to the numerical results obtained for a suitably refined mesh. The modal shapes and the eigenfrequencies approximate accurately the exact solution, as it is shown in Figs. 2–4.

The first eight modes, for each system, are considered. The electric network has been tuned by choosing a proper value of the net inductance, so that the first electric eigenfre-quency matches that of the mechanical system. In this simple case, the tuning procedure is trivial since we have the analytical relations between the net inductance and the electric eigenfrequencies.

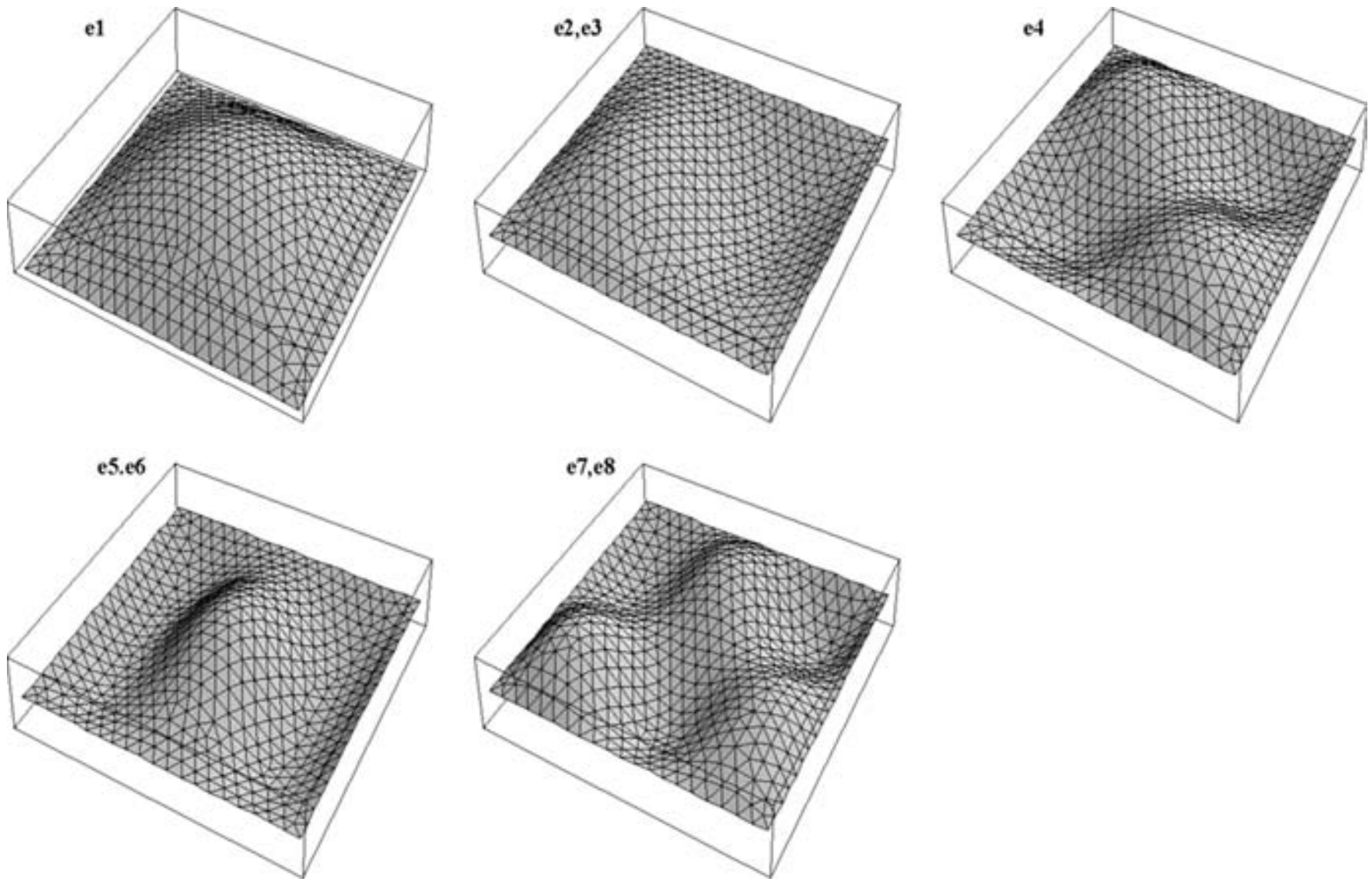

**Fig. 2.** Electric modal shapes

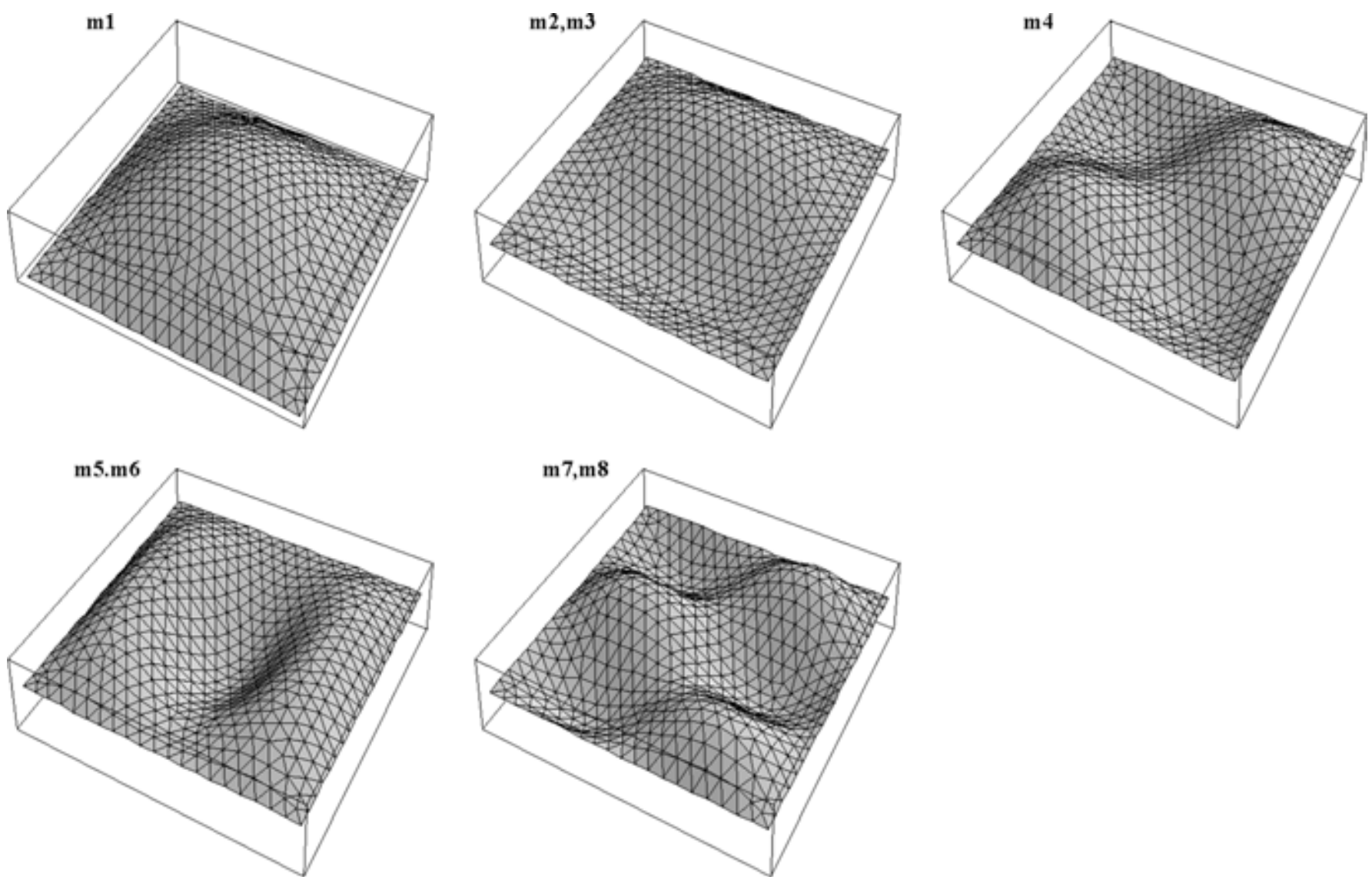

**Fig. 3.** Mechanical modal shapes

| Electric eigenfrequencies | | | Mechanical eigenfrequencies | | |
|---|---|---|---|---|---|
| Numerical results | Exact analytical values | Error (percentage) | Numerical results | Exact analytical values | Error (percentage) |
| 1.002 | 1. | 0.257 | 1.001 | 1. | 0.100 |
| 1.588 | 1.581 | 0.483 | 2.505 | 2.5 | 0.230 |
| 1.588 | 1.581 | 0.491 | 2.506 | 2.5 | 0.263 |
| 2.014 | 2 | 0.721 | 4.015 | 4. | 0.377 |
| 2.255 | 2.236 | 0.864 | 5.023 | 5. | 0.465 |
| 2.255 | 2.236 | 0.883 | 5.024 | 5. | 0.492 |
| 2.577 | 2.549 | 1.102 | 6.536 | 6.5 | 0.562 |
| 2.578 | 2.549 | 1.118 | 6.541 | 6.5 | 0.643 |

**Fig. 4.** Electric and mechanical eigenfrequencies: comparisons with the analytical results

### 4.1.2
### Energy exchange in the coupled system: the undamped case

When considering the coupled system, the results become more interesting. First of all, Fig. 5 shows that the coupling is not negligible only for corresponding modes. The entity of this coupling is due to the spatial inner product of electric and modal shapes. Note that the remarkable coupling shown in Fig. 5 is due to algebraic multiplicity of some eigenvalues.

Let us give a more detailed interpretation to the concept of coupling. Consider a purely mechanical initial condition; for instance, let us assign initial displacements on the first mechanical mode. As it is expected, due to the transducing effect of the PZT actuators, the initial mechanical energy is converted in an electric form, then it returns to the mechanical system, and so on. This effect involves all the energy initially provided to the system; therefore we say that the coupling is at *maximum*. This is due to the fact that the considered mode has been tuned, i.e. it has the same mechanical and electric eigenfrequencies. If we give the second mechanical mode (which is untuned) initial conditions, the amount of energy interchanged is much less than in the previously considered case. Figure 6 shows all these aspects.

Now let's consider the time-evolution of the mechanical and electric displacements. Let's recall that the electric kinematical descriptor, which we call electric displacement, is a time integral of the voltage. Since the corresponding energy is oscillating, we will observe a modulation on the vibrational amplitudes of both systems, with a 90 degrees phase shift between them. In the first considered case (initial conditions on the first mechanical mode), this modulation is actually a complete beating effect. The mechanical vibration, starting from maximum amplitude, decreases to a zero value and than grows up again; vice versa for the electric vibration.

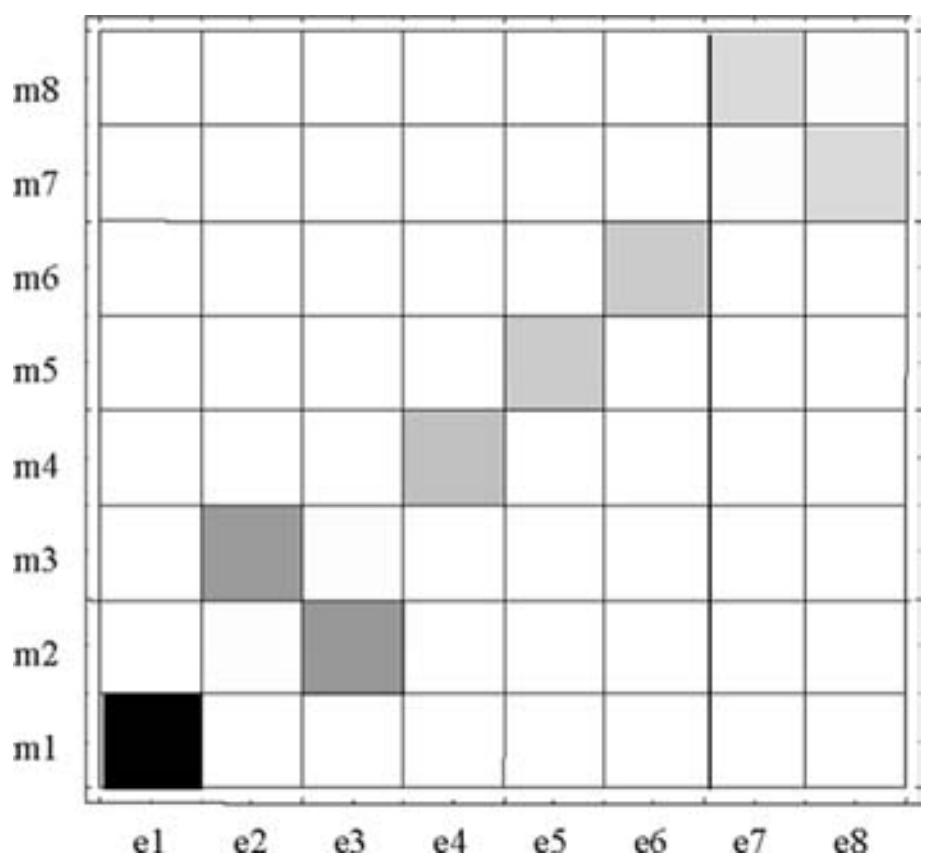

**Fig. 5.** Modal coupling table: back means maximum coupling, white means vanishing coupling

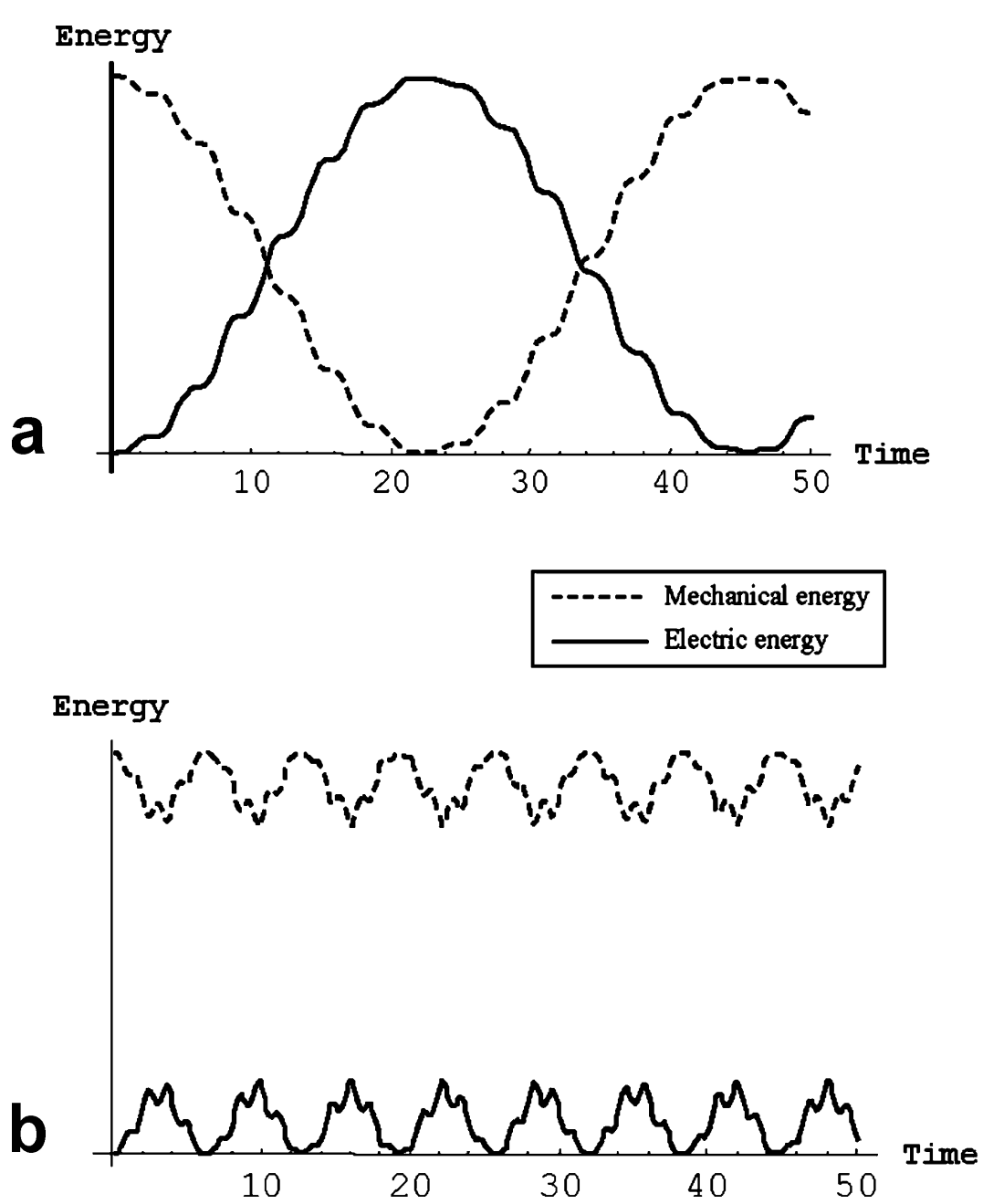

**Fig. 6.** Energy time-evolution for initial conditions on the **a** first mechanical mode; **b** second mechanical mode

### 4.1.3
### Tuned coupled system: damping mechanical vibrations by means of a dissipative network

This is probably the main practical application of this distributed control technique. The idea is the following: if the mechanical energy is converted into the electric form, we could dissipate it with suitable resistors, so that only a fraction of it would return into the mechanical form. After some cycles of energy interchange, the electromechanical would vibrations damp out. Obviously, an optimum value of the network resistance can be found, so that the mechanical damping ratio will be maximized. In other words, using the critical value for the resistance, one can obtain that, once the mechanical energy has been converted into the electric form, the damping is such that this energy will no more return to the mechanical system. According to [17], the critical value for resistance depends on the vibrational mode we want to damp out. Once decided which mode we want to control and damp, both values of network inductance and network resistance are fixed.

Figure 7 shows the time-evolution of the energy of the system, tuned for the first eigenfrequency, starting with the first-mode initial conditions, with the critical damping value suggested in [17].

## 4.2
## Second case: a complex geometry

Following the same steps as in the previous section, let us analyze numerically a more complex case. Let us consider an electromechanical 2-D device with the geometry shown in Fig. 8, with clamped/grounded edges.

We will tune the electric system on the first mechanical mode, that in most of the engineering applications is the critical one. However, with the same procedure it should be possible to tune the electric system on any desired mechanical frequency.

With this example, we will give an idea of the program features, even though much more general cases could be considered. For instance, anisotropic inhomogeneous electromechanical continua could be investigated, as well as any other system whose mathematical model can be arranged in the proper weak formulation. Nevertheless, this example will show how the program can be an useful tool for the design of electromechanical devices.

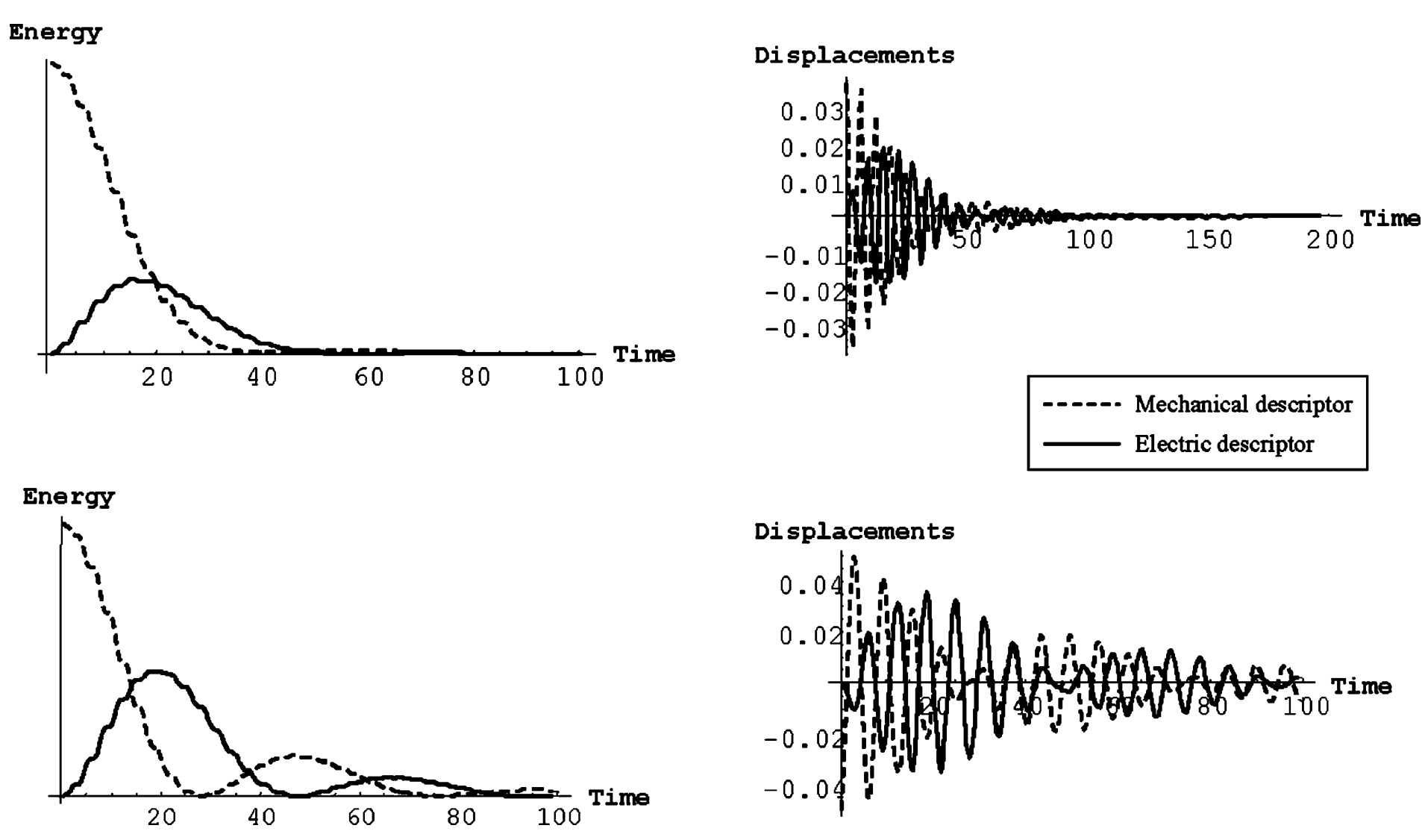

**Fig. 7.** Evolution from purely mechanical initial conditions (first mode): damped case

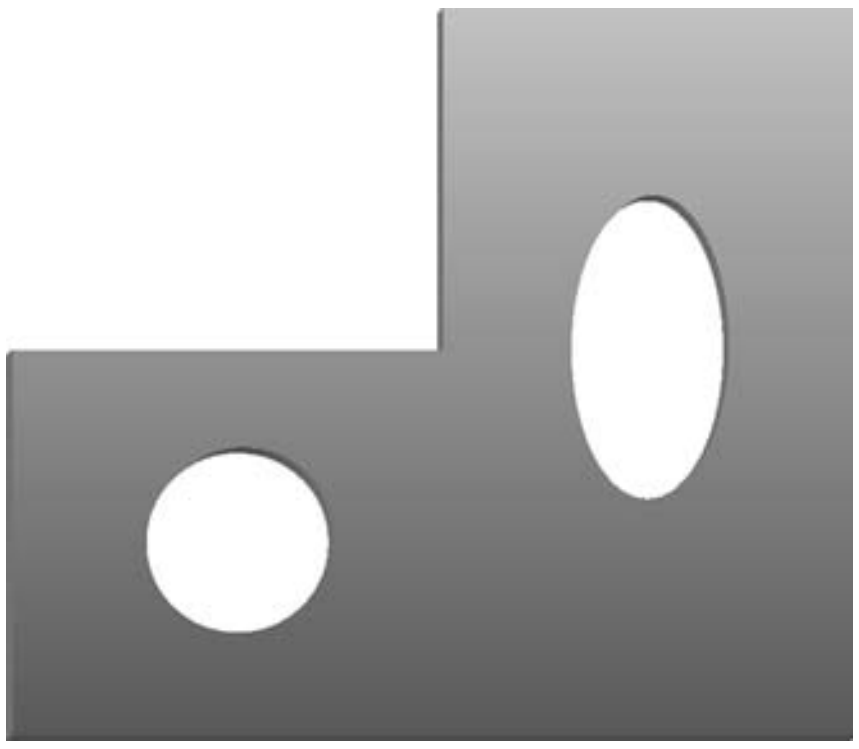

**Fig. 8.** Shape of the device

### 4.2.1
### Uncoupled systems

As it has been described in the previous example, the analysis of the electric and mechanical uncoupled systems is the first step of the process. Here we can verify the accuracy of the tuning procedure, as well as we can check the physical plausibility of the modal shapes. Obviously, in this case we have no analytical solutions to compare the results.

As we can see in Fig. 9, the tuning procedure was efficient, so that, for instance, the first (or the fourth) eigenfrequency is almost the same for both systems. We remark explicitly that in a similar way it is possible to tune any single pair of electric and mechanical frequencies, even if other internal resonances may arise only by chance. It turns out evident that the mechanical eigenfrequencies are now less spaced than in the previous example of the simply supported square plate. This is due to the different geometry and boundary conditions (now the plate is clamped on each edge). As a by-product, this fact has the consequence that also other eigenfrequencies can happen to be almost tuned. Therefore we may expect better coupling performances from this device.

Furthermore, the modal shapes of the electric and mechanical systems differ much more because of the clamped edges: we cannot impose zero slopes on the boundaries for a second order system as the electric one, viz. Figs. 10 and 11.

Hence the coupling will be probably more complicated; that is, we expect each electric mode to be coupled with more than one mechanical mode, and viceversa.

In fact, as deduced from Fig. 12, we note that there is a distributed modal coupling: the matrix of inner products between electric and mechanical eigenvectors is not anymore diagonal. This fact has important consequences in the dynamics of the whole system. In fact, if we supply energy to a given mechanical mode, this will couple with a large number of electric modes, so that, once the energy returns in the mechanical form, a multi-modal motion appears. However, this effect is limited by the fact that only tuned modes have a complete coupling.

### 4.2.2
### Tuned coupled system: coupling performances and the placement of a mechanical impulse

As we expected from the previous analysis, now there is a good coupling in the first few modes, since the first eigenfrequencies are close enough. This fact will have some important consequences in the performances of the system when used as a vibration damper.

Let us firstly consider the case of purely mechanical uni-modal initial conditions, Fig 13. Again, as a by-product, we find that also the third eigenfrequency comes out to be almost tuned, so that we have a very good energy interchange when exciting the third mechanical mode. The second mode, for the same reason, presents an appreciable coupling yet, but starting from the fourth mode the coupling performances decrease rapidly since the eigenfrequencies of the electric and mechanical subsystems become very different.

Now let us consider a mechanical impulse localized in a point of the structure. Projecting on the chosen base of uncoupled modes, we get a superposition of all the considered modes. The coefficients of such superposition depend upon the point of application of the impulse. Therefore, also the coupling performances are a function of the impulse placement.

Figure 14 shows this concept. In the first case, the impulse is localized in a central area of the plate, so that we excite mostly the first mechanical mode, which is efficiently coupled. Hence we get a good ratio of energy interchange. Instead, if the impulse is applied in a peripheral area of the plate, as it is shown in the second column of Fig. 14, we excite higher modes which are not tuned and therefore not efficiently coupled. This fact implies that a smaller part of mechanical energy is converted into the electric form. However, the performances are still acceptable.

### 4.2.3
### Coupled system: damping mechanical vibrations by means of a dissipative network

If we add some damping to the electric network, the whole system will not be conservative anymore. Hence, this could be a way to dissipate the mechanical energy once converted into the electric form. However, as we noticed in the square plate case, the damping ratio must be maximized by choosing the optimum value for the network resistance, otherwise the damping performances of the system would be unacceptable. While the analytical studies in [17] provide that optimum value for the simply supported square plate, in general cases we have to numerically explore the system to get an approximate value for the optimal resistance. This is exactly what is done here. For this analysis, we consider the case of tuning on the first mode, which seems to have better performances. Performing numerical experiments, we can estimate in an iterative fashion the maximum mechanical damping ratio, as it is shown in Fig. 15, where we present the cases of sub-critical, critical and super-critical damping values. Let us note that, if the numerical value of the resistance is too large, the coupling vanishes and, while the electric vibrations decrease rapidly, the mechanical ones present a very low damping ratio.

| 1° MODE TUNED | | 4° MODE TUNED | |
|---|---|---|---|
| Electric eigenfrequencies | Mechanical eigenfrequencies | Electric eigenfrequencies | Mechanical eigenfrequencies |
| **1.101** | **1.101** | 1.398 | 1.101 |
| 1.445 | 1.513 | 1.832 | 1.513 |
| 1.598 | 1.620 | 2.026 | 1.620 |
| 1.835 | 2.276 | **2.276** | **2.276** |
| 2.000 | 2.490 | 2.548 | 2.490 |
| 2.031 | 2.691 | 2.598 | 2.691 |
| 2.160 | 3.057 | 2.765 | 3.057 |
| 2.172 | 3.141 | 2.790 | 3.141 |
| 2.413 | 3.354 | 3.101 | 3.354 |

**Fig. 9.** Table of eigenfrequencies

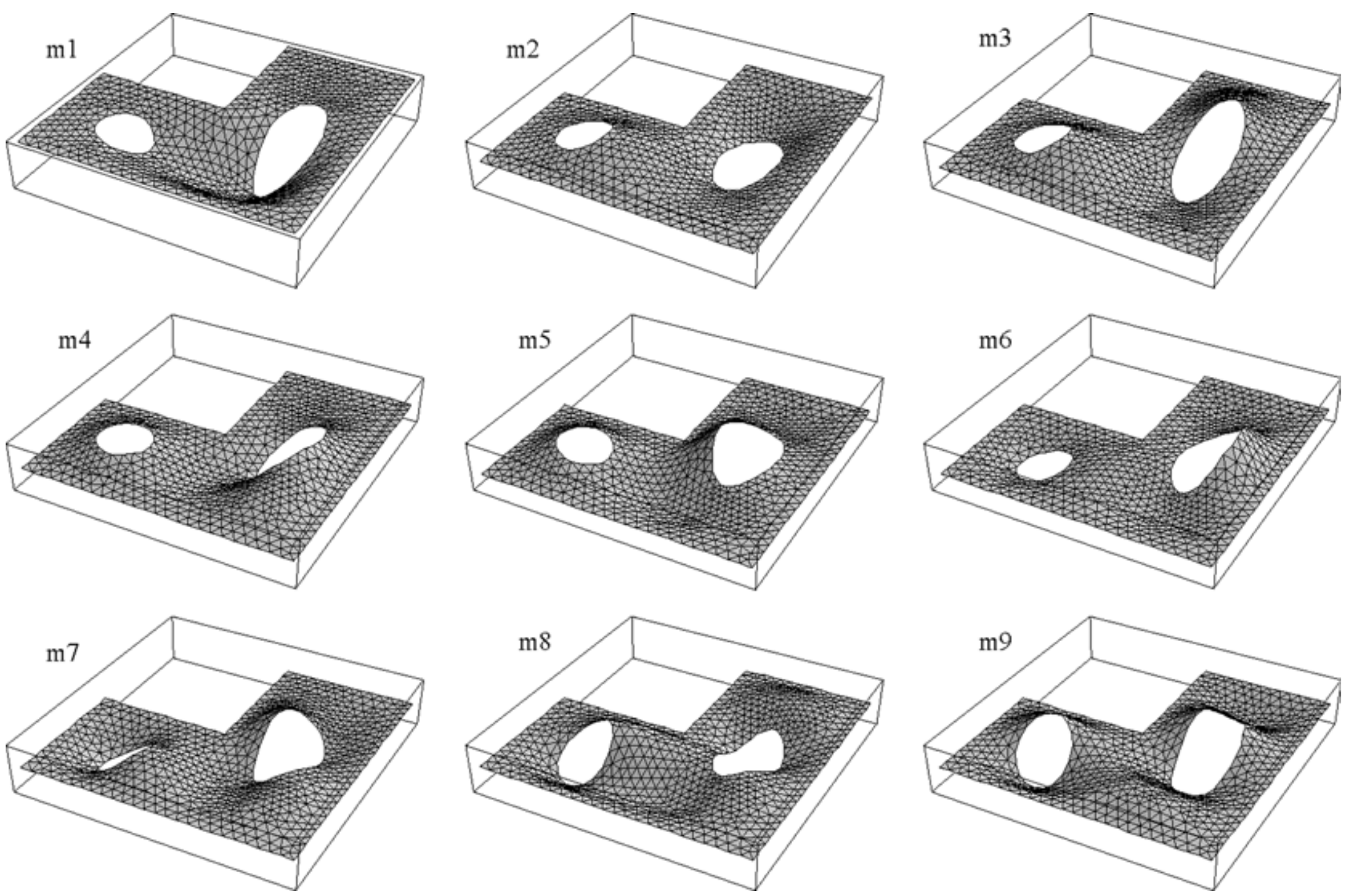

**Fig. 10.** Mechanical modal shapes

Once the optimum value for the network resistance has been determined, we can analyze the overall damping performances with the usual set of initial conditions: unimodal (first four modes), Fig. 16, and impulsive (two different impulse placements), Fig. 17.

The numerical study of this device leads to some useful results from an engineering point of view. We can summarize them in the following two main areas:

1. We have obtained the optimum network parameters, which are necessary in designing such devices;
2. We have performed an accurate analysis of the system dynamics, which allows us to forecast its behavior and its performances.

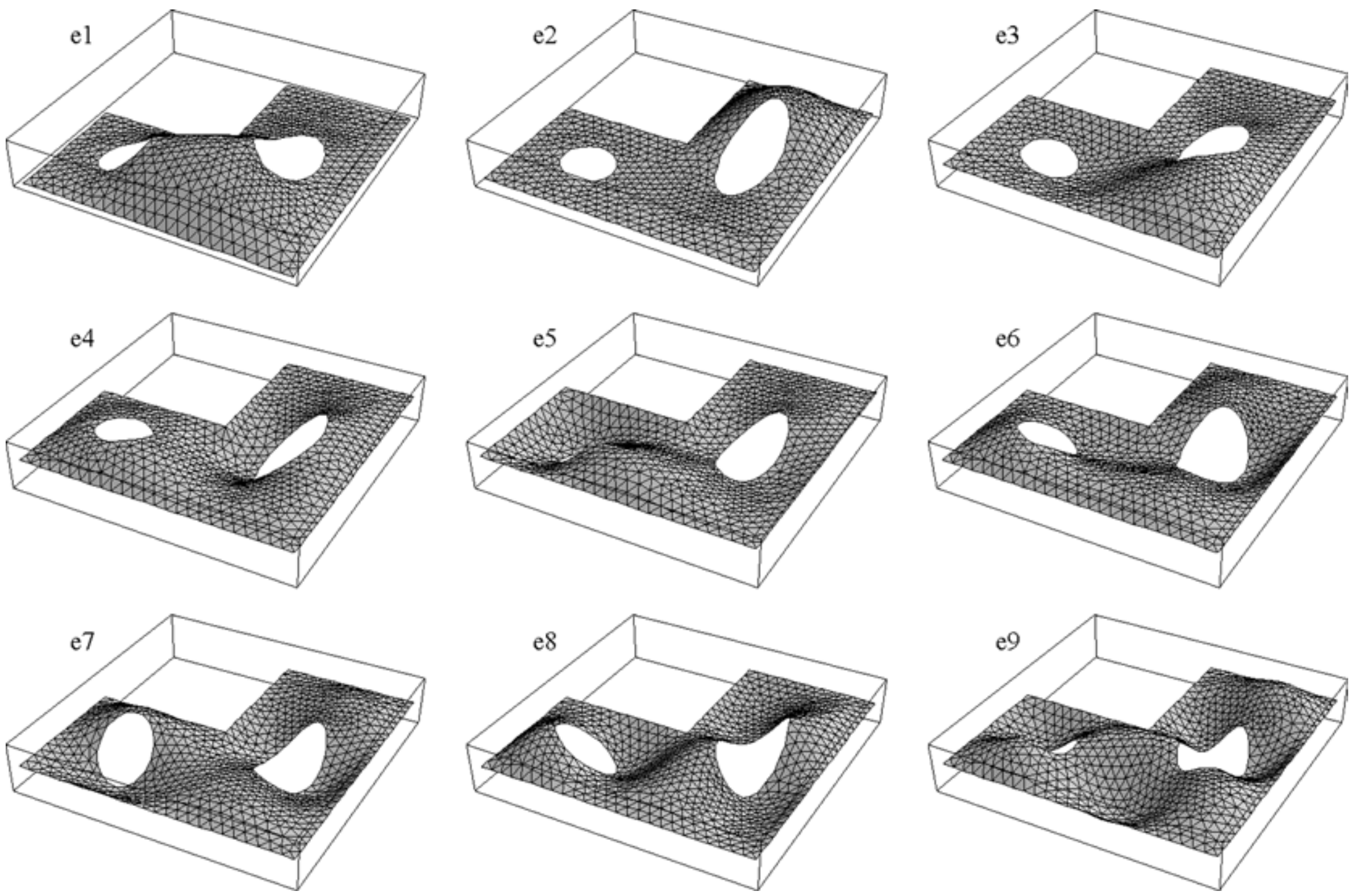

**Fig. 11.** Electric modal shapes

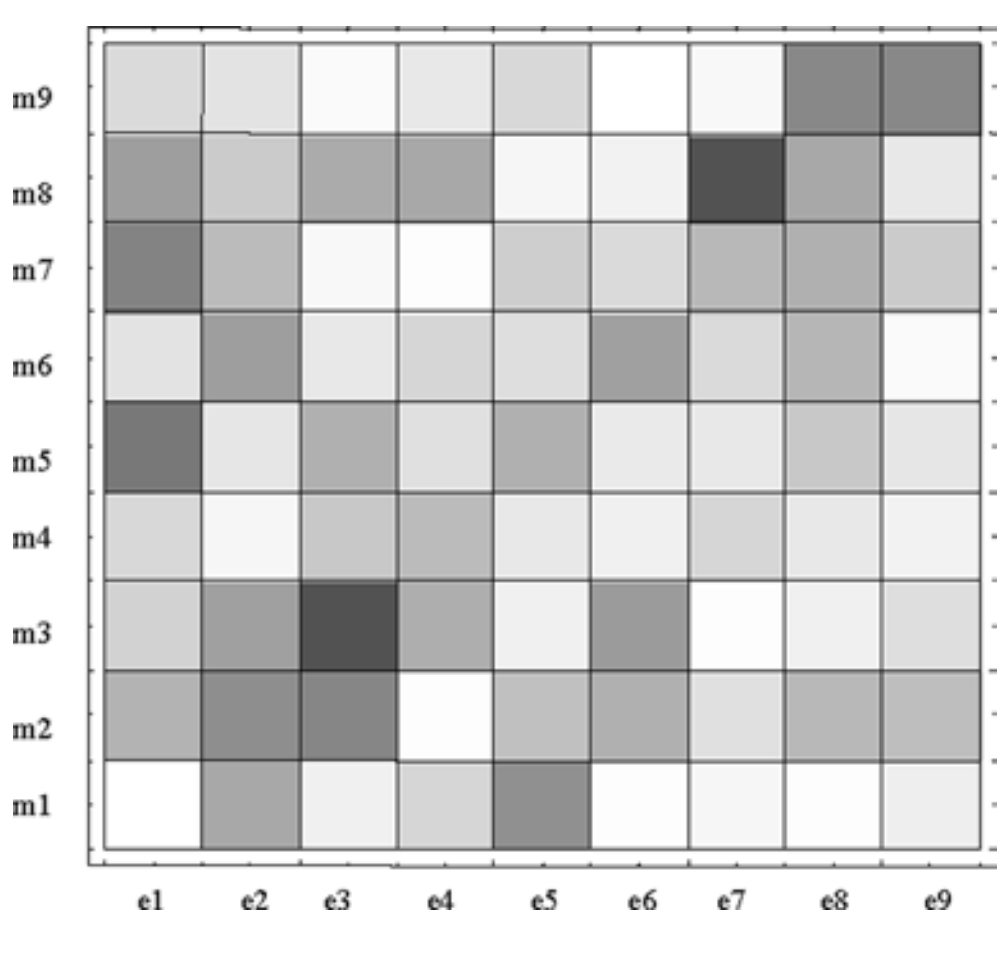

**Fig. 12.** Table of modal coupling, s. Fig. 5

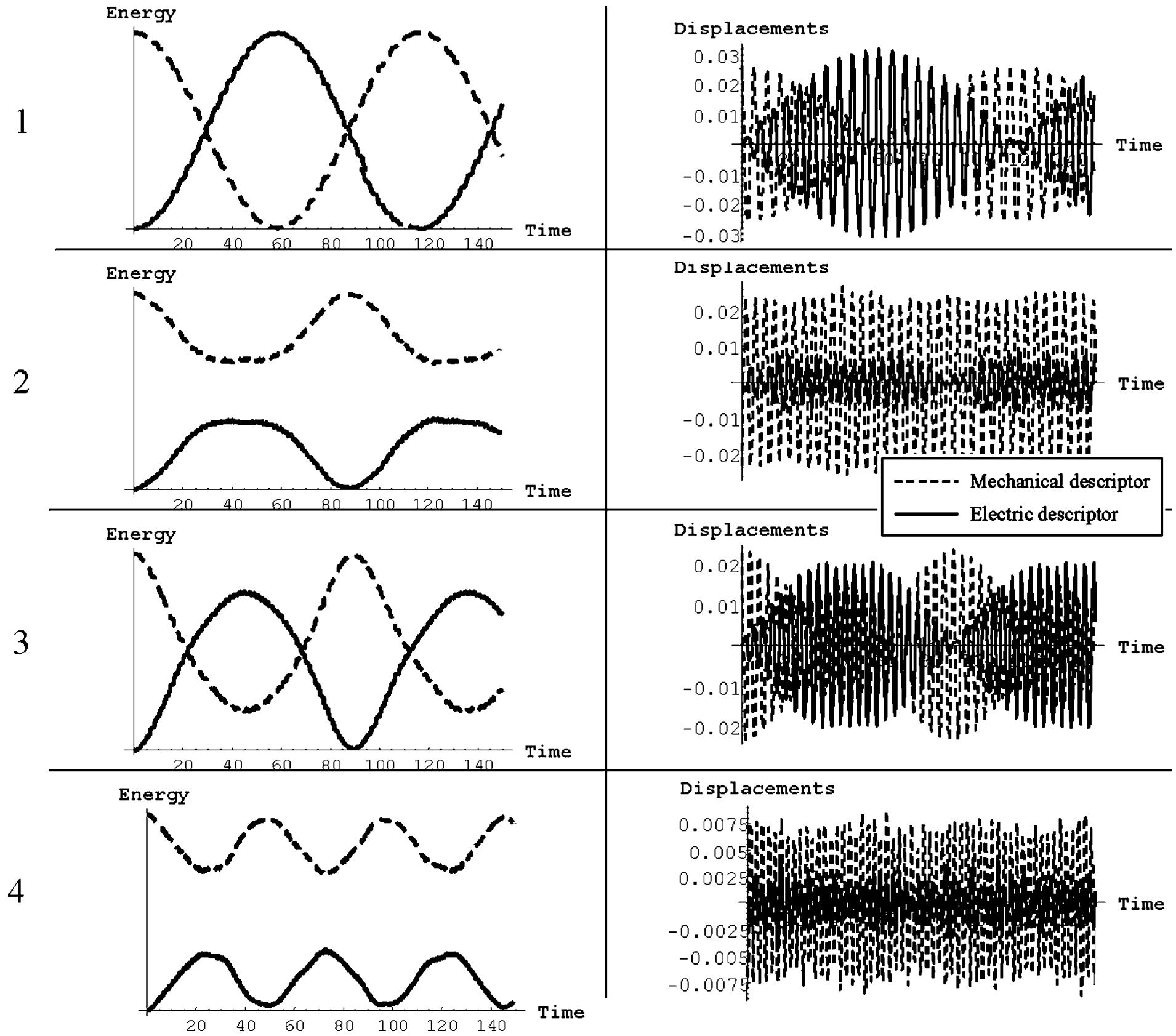

**Fig. 13.** Time-evolution from different uni-modal initial conditions

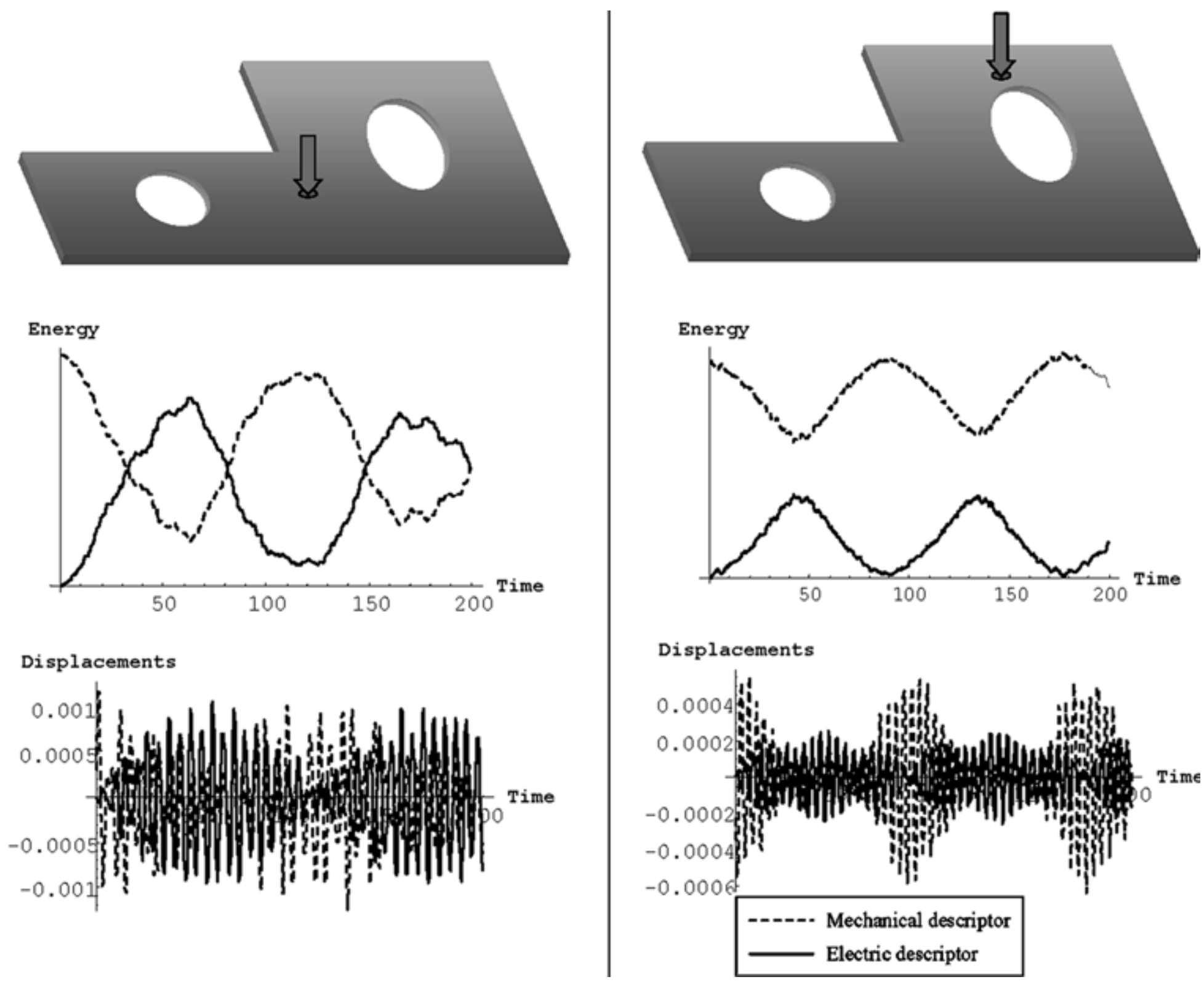

**Fig. 14.** Time-evolution from different impulsive initial conditions

## 5
## Concluding remarks

A new FE code has been realized. Its wide range of applications is probably its best feature, as it allows for the analysis of coupled systems of second- and fourth-order PDEs with dissipative terms. The usage of the code is simple and user-oriented, with few control parameters to be set to control the accuracy and its main aspects. As many procedures as possible have been made completely automatic, so that changing some input parameter is easy and fast. All those characteristics have a price, which is the CPU-time needed to perform the computations. Obviously, a dedicated software, realized to solve only a class of particular problems, can easily be optimized and it can reach a higher speed requiring less hardware power. Nevertheless, systems involving two or more PDEs of different nature coupled in various fashions require a very flexible computational method.

The program has been applied to the study of PEM plates, [17], in which vibration control is obtained by means of piezoelectric actuation. A weak formulation of the evolution equation of such systems has been derived and the data input needed by the code have been computed. A test case, e.g. a simply supported square PEM plate, has been used to prove the algorithm correctness and the results accuracy. Subsequently, a more general case has been investigated, and the overall performances of the system have been shown. The code has proven to be useful in the design of such electromechanical systems, since it allowed us to compute the electric parameters needed to optimize the mechanical vibration control and damping.

The obtained results show that (i) piezoelectric actuators, interconnected by means of simple passive electric networks, can control mechanical vibration in a fixed frequency range and (ii) more sophisticated electronic circuits could be similarly efficient in a broader band of frequencies.

## Appendix: Some details on the numerical code

### A.1
### The Galerkin formulation

Some details are given on the numerical procedure to obtain an approximate solution of Problem 1.

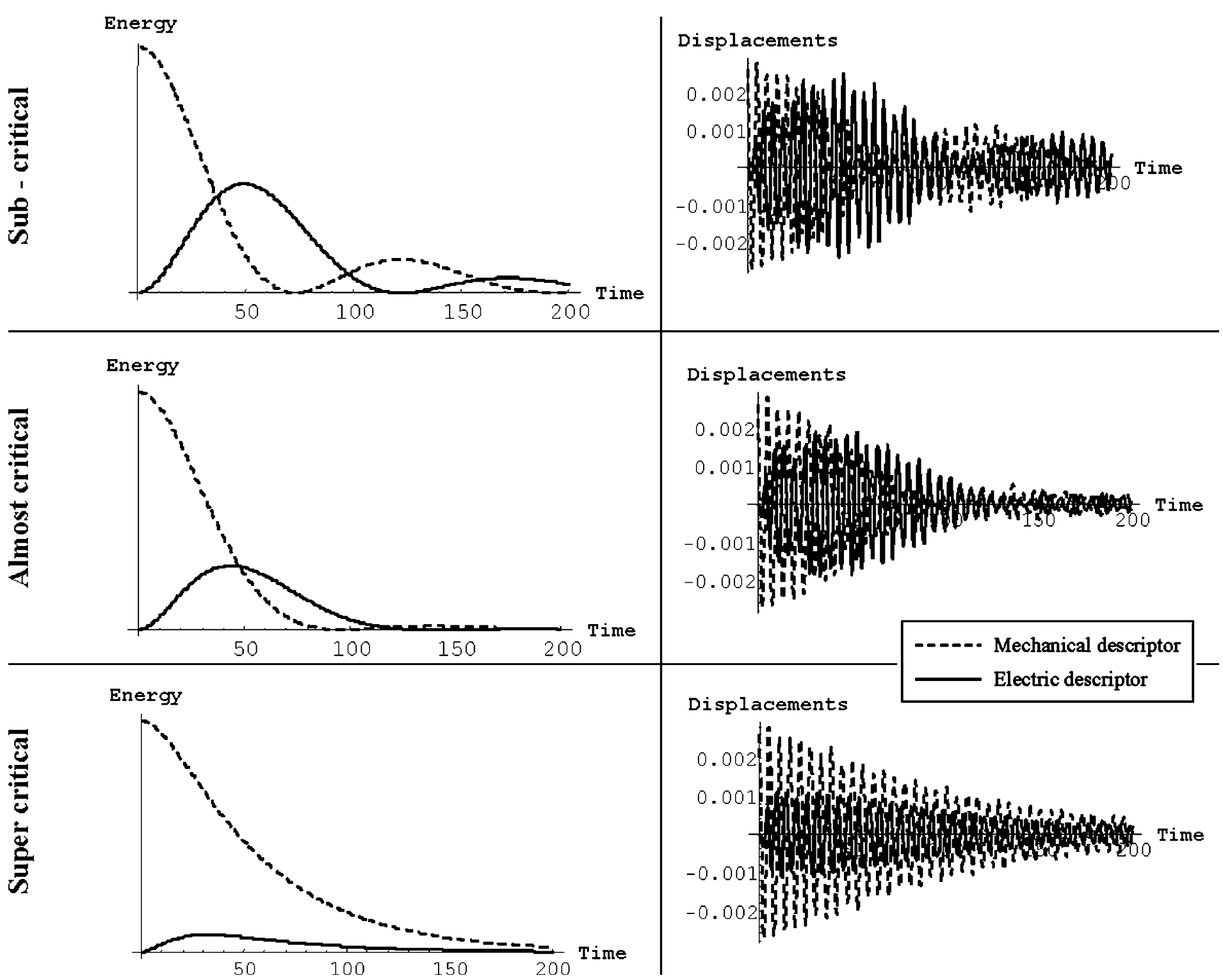

**Fig. 15.** Time-evolution from first mode initial conditions: three different damping ratios

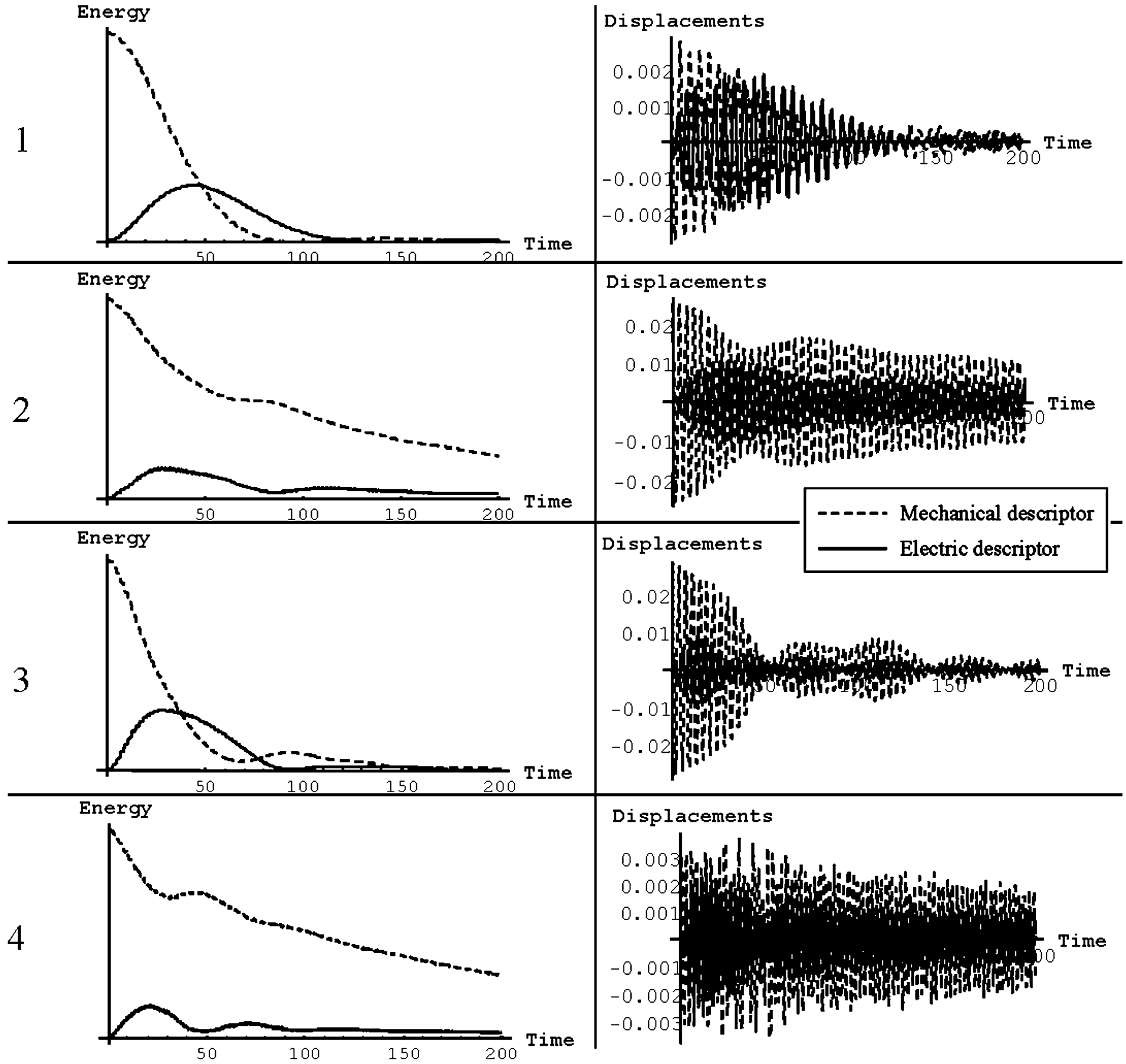

**Fig. 16.** Time-evolution from uni-modal initial conditions: damped case, first four modes

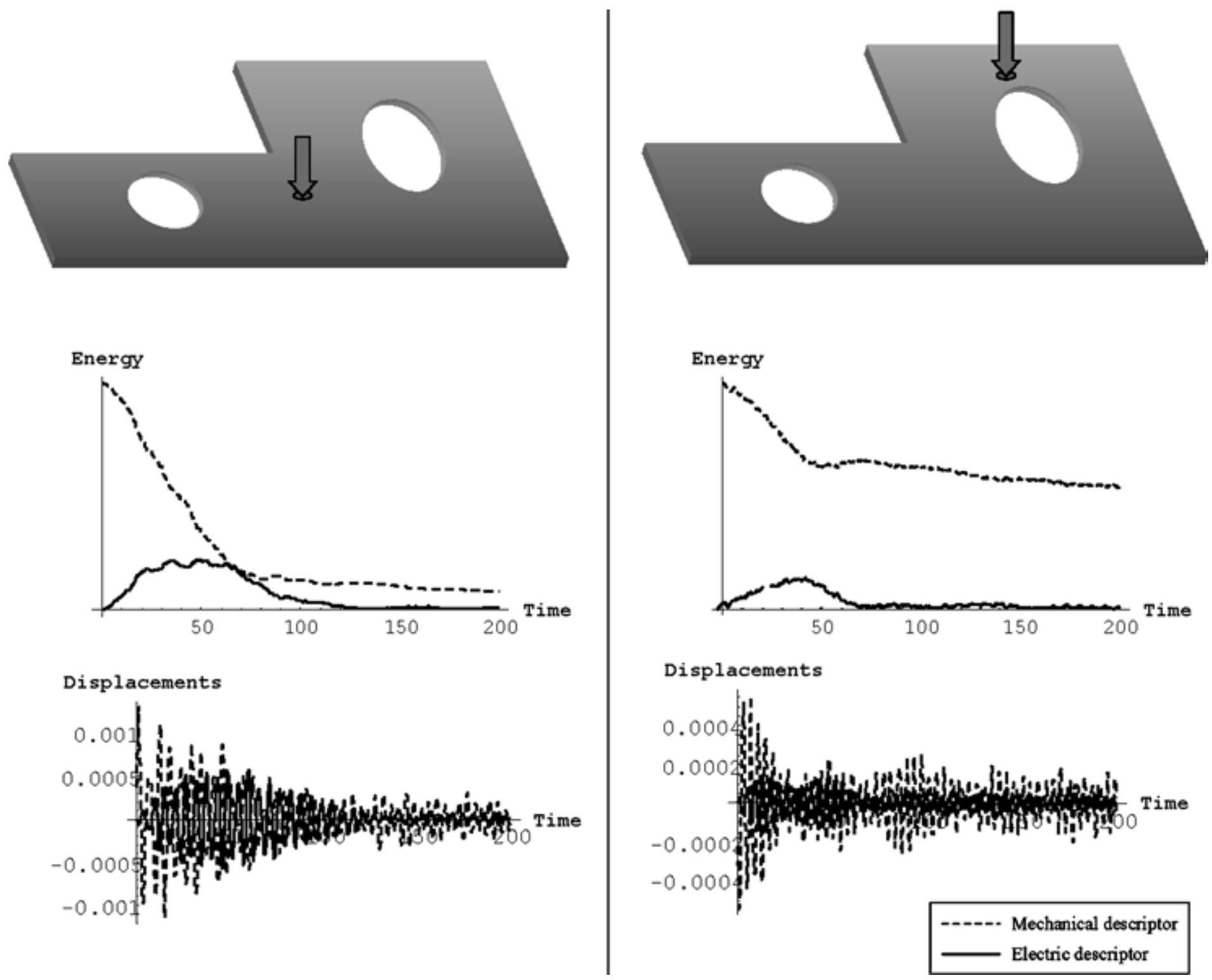

**Fig. 17.** Time-evolution from different impulsive initial conditions: damped case

Substituting the constitutive relations in the virtual work principle we obtain

$$\begin{aligned}&\langle \mathbf{G}\ddot{\mathbf{u}} + \mathbf{S}\dot{\mathbf{u}} + \mathbf{T}\mathbf{u} + \mathbf{V}\dot{\boldsymbol{\varepsilon}} + \mathbf{b}_0, \dot{\mathbf{u}}^t\rangle_s + \langle \mathbf{G}_B^1\ddot{\mathbf{u}}_{,1} + \mathbf{B}_1^0, \dot{\mathbf{u}}_{,1}^t\rangle_s + \langle \mathbf{G}_B^2\ddot{\mathbf{u}}_{,2} + \mathbf{B}_2^0, \dot{\mathbf{u}}_{,2}^t\rangle \\ &\quad + \langle \mathbf{f}_0, \dot{\mathbf{u}}^t\rangle_{\partial s} + \langle \mathbf{s}_1^0, \dot{\mathbf{u}}_{,1}^t\rangle_{\partial s} + \langle \mathbf{s}_2^0, \dot{\mathbf{u}}_{,2}^t\rangle_{\partial s} - \langle \mathbf{E}\boldsymbol{\varepsilon} + \mathbf{C}\dot{\mathbf{u}} + \mathbf{R}\mathbf{u} + \boldsymbol{\sigma}_0, \dot{\boldsymbol{\varepsilon}}^t\rangle_{\mathcal{S}} = 0 \ . \end{aligned} \tag{A1}$$

Following the steps of a standard FE procedure, subdivide the domain $\mathcal{S}$ into subdomains $\mathcal{D}^e$: $\mathcal{D}^1 \cup \mathcal{D}^2 \cup \cdots \cup \mathcal{D}^{ne} = \mathcal{S}$ and $\mathcal{D}^i \cap \mathcal{D}^j = \emptyset$ for $i \neq j$. Hence the integrals in (A1) can be expressed as summations of integrals computed on each element $\mathcal{D}^e$,

$$\begin{aligned}&\sum_e \langle \mathbf{G}\ddot{\mathbf{u}} + \mathbf{S}\dot{\mathbf{u}} + \mathbf{T}\mathbf{u} + \mathbf{V}\dot{\boldsymbol{\varepsilon}} + \mathbf{b}_0, \dot{\mathbf{u}}^t\rangle_{\mathcal{D}^e} + \sum_e \langle \mathbf{G}_B^1\ddot{\mathbf{u}}_{,1} + \mathbf{B}_1^0, \dot{\mathbf{u}}_{,1}^t\rangle_{\mathcal{D}^e} + \sum_e \langle \mathbf{G}_B^2\ddot{\mathbf{u}}_{,2} + \mathbf{B}_2^0, \dot{\mathbf{u}}_{,2}^t\rangle \\ &\quad + \sum_e \langle \mathbf{f}_0, \dot{\mathbf{u}}^t\rangle_{\partial\mathcal{D}^e\cap\partial s} + \sum_e \langle \mathbf{s}_1^0, \dot{\mathbf{u}}_{,1}^t\rangle_{\partial\mathcal{D}^e\cap\partial s} + \sum_e \langle \mathbf{s}_2^0, \dot{\mathbf{u}}_{,2}^t\rangle_{\partial\mathcal{D}^e\cap\partial s} - \sum_e \langle \mathbf{E}\boldsymbol{\varepsilon} + \mathbf{C}\dot{\mathbf{u}} + \mathbf{R}\mathbf{u} + \boldsymbol{\sigma}_0, \dot{\boldsymbol{\varepsilon}}^t\rangle_{\mathcal{D}^e} = 0 \ .\end{aligned}$$

Consider a single element. The displacements vector can be locally expressed in terms of basis function defined only in the element

$$\mathbf{u}^e(\mathbf{r}, t) = \mathbf{N}^e(\mathbf{r})\mathbf{q}^e(t) \tag{A2}$$

where $\mathbf{N}^e(\mathbf{r})$ is the shape functions matrix, while $\mathbf{q}(t)$ is the vector of the nodal values of $\mathbf{u}^e(\mathbf{r}, \mathrm{t})$.
Let us define

$$\boldsymbol{\varepsilon}^e = \mathbf{N}_\varepsilon^e \mathbf{q}^e \ , \tag{A3}$$

$$\boldsymbol{\varepsilon}^{et} = \mathbf{N}_\varepsilon^{et} \mathbf{q}^{et} \ , \tag{A4}$$

where $\mathbf{N}_\varepsilon^e$ and $\mathbf{N}_\varepsilon^{et}$ are given by the following linear combination:

$$\mathbf{N}_\varepsilon^e = \mathbf{H}_0\mathbf{N}^e + \mathbf{H}_1\mathbf{N}_{,1}^e + \mathbf{H}_2\mathbf{N}_{,2}^e + \mathbf{H}_3\mathbf{N}_{,11}^e + \mathbf{H}_4\mathbf{N}_{,22}^e + \mathbf{H}_5\mathbf{N}_{,12}^e \ , \tag{A5}$$

$$\mathbf{N}_\varepsilon^{et} = \mathbf{H}_0^t\mathbf{N}^e + \mathbf{H}_1^t\mathbf{N}_{,1}^e + \mathbf{H}_2^t\mathbf{N}_{,2}^e + \mathbf{H}_3^t\mathbf{N}_{,11}^e + \mathbf{H}_4^t\mathbf{N}_{,22}^e + \mathbf{H}_5^t\mathbf{N}_{,12}^e \ . \tag{A6}$$

The contribution of a single element to the integral formulation is

$$\begin{aligned}&\ddot{\mathbf{q}}^{eT}\langle \mathbf{G}\mathbf{N}^e, \mathbf{N}^e\rangle_{\mathcal{D}^e}\dot{\mathbf{q}}^{et} + \dot{\mathbf{q}}^{eT}\langle \mathbf{S}\mathbf{N}^e, \mathbf{N}^e\rangle_{\mathcal{D}^e}\dot{\mathbf{q}}^{et} + \mathbf{q}^{eT}\langle \mathbf{T}\mathbf{N}^e, \mathbf{N}^e\rangle_{\mathcal{D}^e}\dot{\mathbf{q}}^{et} + \dot{\mathbf{q}}^{eT}\langle \mathbf{V}\mathbf{N}_\varepsilon^e, \mathbf{N}^e\rangle_{\mathcal{D}^e}\dot{\mathbf{q}}^{et} \\ &\quad + \langle \mathbf{b}_0, \mathbf{N}^e\rangle_{\mathcal{D}^e}\dot{\mathbf{q}}^{et} + \ddot{\mathbf{q}}^{eT}\langle \mathbf{G}_B^1\mathbf{N}_{,1}^e, \mathbf{N}_{,1}^e\rangle_{\mathcal{D}^e}\dot{\mathbf{q}}^{et} + \ddot{\mathbf{q}}^{eT}\langle \mathbf{G}_B^2\mathbf{N}_{,2}^e, \mathbf{N}_{,2}^e\rangle_{\mathcal{D}^e}\dot{\mathbf{q}}^{et} + \langle \mathbf{B}_1^0, \mathbf{N}_{,1}^e\rangle_{\mathcal{D}^e}\dot{\mathbf{q}}^{et} \\ &\quad + \langle \mathbf{B}_2^0, \mathbf{N}_{,2}^e\rangle_{\mathcal{D}^e}\dot{\mathbf{q}}^{et} + \langle \mathbf{f}_0, \mathbf{N}^e\rangle_{\partial\mathcal{D}^e\cap\partial\int}\dot{\mathbf{q}}^{et} + \langle \mathbf{s}_1^0, \mathbf{N}_{,1}^e\rangle_{\partial\mathcal{D}^e\cap\partial\int}\dot{\mathbf{q}}^{et} + \langle \mathbf{s}_2^0, \mathbf{N}_{,2}^e\rangle_{\partial\mathcal{D}^e\cap\partial\int}\dot{\mathbf{q}}^{et} \\ &\quad - \mathbf{q}^{eT}\langle \mathbf{E}\mathbf{N}_\varepsilon^e, \mathbf{N}_\varepsilon^{et}\rangle_{\mathcal{D}^e}\dot{\mathbf{q}}^{et} - \dot{\mathbf{q}}^{eT}\langle \mathbf{C}\mathbf{N}^e, \mathbf{N}_\varepsilon^{et}\rangle_{\mathcal{D}^e}\dot{\mathbf{q}}^{et} - \dot{\mathbf{q}}^{eT}\langle \mathbf{R}\mathbf{N}^e, \mathbf{N}_\varepsilon^{et}\rangle_{\mathcal{D}^e}\dot{\mathbf{q}}^{et} - \langle \boldsymbol{\sigma}_0, \mathbf{N}_\varepsilon^{et}\rangle_{\mathcal{D}^e}\dot{\mathbf{q}}^{et}\end{aligned}$$

where $T$ holds for *transpose*. Grouping the terms we get

$$\begin{aligned}&\ddot{\mathbf{q}}^{eT}(\langle \mathbf{G}\mathbf{N}^e, \mathbf{N}^e\rangle_{\mathcal{D}^e} + \langle \mathbf{G}_B^1\mathbf{N}_{,1}^e, \mathbf{N}_{,1}^e\rangle_{\mathcal{D}^e} + \langle \mathbf{G}_B^2\mathbf{N}_{,2}^e, \mathbf{N}_{,2}^e\rangle_{\mathcal{D}^e})\dot{\mathbf{q}}^{et} + \dot{\mathbf{q}}^{eT}(\langle \mathbf{S}\mathbf{N}^e, \mathbf{N}^e\rangle_{\mathcal{D}^e} \\ &\quad + \langle \mathbf{V}\mathbf{N}_\varepsilon^e, \mathbf{N}^e\rangle_{\mathcal{D}^e} - \langle \mathbf{C}\mathbf{N}^e, \mathbf{N}_\varepsilon^{et}\rangle_{\mathcal{D}^e})\dot{\mathbf{q}}^{et} + \mathbf{q}^{eT}(\langle \mathbf{T}\mathbf{N}^e, \mathbf{N}^e\rangle_{\mathcal{D}^e} - \langle \mathbf{E}\mathbf{N}^e\varepsilon, \mathbf{N}_\varepsilon^{et}\rangle_{\mathcal{D}^e} \\ &\quad - \langle \mathbf{R}\mathbf{N}^e, \mathbf{N}_\epsilon^{et}\rangle_{\mathcal{D}^e})\dot{\mathbf{q}}^{et} + (\langle \mathbf{b}_0, \mathbf{N}^e\rangle_{\mathcal{D}^e} + \langle \mathbf{B}_1^0, \mathbf{N}_{,1}^e\rangle_{\mathcal{D}^e} + \langle \mathbf{B}_2^0, \mathbf{N}_{,2}^e\rangle_{\mathcal{D}^e} \\ &\quad + \langle \mathbf{f}_0, \mathbf{N}^e\rangle_{\partial\mathcal{D}^e\cap\partial\int} + \langle \mathbf{s}_1^0, \mathbf{N}_{,1}^e\rangle_{\partial\mathcal{D}^e\cap\partial\int} + \langle \mathbf{s}_2^0, \mathbf{N}_{,2}^e\rangle_{\partial\mathcal{D}^e\cap\partial\mathcal{S}} - \langle \boldsymbol{\sigma}_0, \mathbf{N}_\varepsilon^{et}\rangle_{\mathcal{D}^e})\dot{\mathbf{q}}^{et}\end{aligned}$$

Hence we can define:

| | |
|---|---|
| Local mass matrix | $\mathbf{K}_2^e = \langle \mathbf{G}\mathbf{N}^e, \mathbf{N}^e\rangle_{\mathcal{D}^e} + \langle \mathbf{G}_B^1, \mathbf{N}_{,1}^e, \mathbf{N}_{,1}^e\rangle_{\mathcal{D}^e} + \langle \mathbf{G}_B^2, \mathbf{N}_{,2}^e, \mathbf{N}_{,2}^e\rangle_{\mathcal{D}^e}$ |
| Local dampling/coupling matrix | $\mathbf{K}_1^e = \langle \mathbf{S}\mathbf{N}^e, \mathbf{N}^e\rangle_{\mathcal{D}^e} + \langle \mathbf{V}\mathbf{N}_\varepsilon^e, \mathbf{N}^e\rangle_{\mathcal{D}^e} - \langle \mathbf{C}\mathbf{N}^e, \mathbf{N}_\varepsilon^{et}\rangle_{\mathcal{D}^e}$ |
| Local stiffness matrix | $\mathbf{K}_0^e = \langle \mathbf{T}\mathbf{N}^e, \mathbf{N}^e\rangle_{\mathcal{D}^e} - \langle \mathbf{E}\mathbf{N}_\varepsilon^e, \mathbf{N}_\varepsilon^{et}\rangle_{\mathcal{D}^e} - \langle \mathbf{R}\mathbf{N}^e, \mathbf{N}_\varepsilon^{et}\rangle_{\mathcal{D}^e}$ |
| Local load vector | $F^e = -\langle \mathbf{b}_0, \mathbf{N}^e\rangle_{\mathcal{D}^e} - \langle \mathbf{B}_1^0, \mathbf{N}_{,1}^e\rangle_{\mathcal{D}^e} - \langle \mathbf{B}_2^0, \mathbf{N}_{,2}^e\rangle_{\mathcal{D}^e} - \langle \mathbf{f}_0, \mathbf{N}^e\rangle_{\partial\mathcal{D}^e\cap\partial\mathcal{S}}$ $- \langle \mathbf{s}_1^0, \mathbf{N}_{,1}^e\rangle_{\partial\mathcal{D}^e\cap\partial\mathcal{S}} - \langle \mathbf{s}_2^0, \mathbf{N}_{,2}^e\rangle_{\partial\mathcal{D}^e\cap\partial\mathcal{S}} + \langle \sigma_0, \mathbf{N}_\varepsilon^{et}\rangle_{\mathcal{D}^e}$ |

(A7)

The next step is to assemble the local matrices and vector to get the final system. Following standard FE procedures, based upon the correspondence between local and global node numbering, we finally obtain

$$\mathbf{K}_2\ddot{\mathbf{q}}(t) + \mathbf{K}_1\dot{\mathbf{q}}(t) + \mathbf{K}_0\mathbf{q}(t) = \mathbf{F}(t) \tag{A8}$$

where $\mathbf{K}_2, \mathbf{K}_1, \mathbf{K}_0$ are respectively the global mass, damping/coupling and stiffness matrices, $\mathbf{F}$ is the global load vector and $\mathbf{q}$ is the vector of nodal generalized displacements.

Note that the dimensions of this system of equations can be very large; a quiet refined mesh of the domain can easily reach thousands of nodes, so that a numerical analysis such system would be too heavy. Therefore, a different approach is applied. First we perform a modal analysis of the system, assuming $\mathbf{K}_1$ to be small compared to $\mathbf{K}_0$ and $\mathbf{K}_2$; then we project all the matrices involved over a base of a suitable number of eigenvectors. In this way we get a reduced system of much smaller dimensions. Obviously, also the initial conditions have to be projected on the chosen base, loosing something in terms of accuracy in reproducing Dirac-delta like functions, but gaining much more in terms of computational time. This procedure can be illustrated as follows. Let us define

$$\mathbf{q}(t) = \mathbf{a}e^{jwt} \quad ,$$

so that the homogeneous problem, neglecting $\mathbf{K}_1$, reads as follows:

$$-\omega^2\mathbf{k}_2\mathbf{a} + \mathbf{K}_0\mathbf{a} = \mathbf{0} \quad ,$$

which, with some manipulation, becomes:

$$\mathbf{K}_0^{-1}\mathbf{K}_2\mathbf{a} = \frac{1}{\omega^2}\mathbf{a} \quad .$$

Hence the eigensystem of $\mathbf{K}_0^{-1}\mathbf{K}_2$ will give us eigenvectors and eigenfrequency of the uncoupled/undamped system. Let us define the $i$-th eigenvector $\mathbf{a}_i$, and let us choose a number $n$ of them to build up an orthonormal base. Now let us define:

$$\mathbf{T} = \begin{bmatrix} \mathbf{a}_{11} & \mathbf{a}_{12} & \dots & \mathbf{a}_{1m} \\ \mathbf{a}_{21} & \dots & \dots & \dots \\ \dots & \dots & \dots & \dots \\ \dots & \dots & \dots & \dots \\ \mathbf{a}_{n1} & \mathbf{a}_{n2} & \dots & \mathbf{a}_{nm} \end{bmatrix} \quad ,$$

where $m$ is the dimension of the initial system of equation. Now let us consider the projection of the nodal displacements vector $\mathbf{q}$ on this base:

$$\begin{aligned} \mathbf{z} &= \mathbf{T}\mathbf{q}, \\ \mathbf{q} &= \mathbf{T}^T\mathbf{Z} \quad . \end{aligned} \tag{A9}$$

Therefore, substituting this expansion in (A8) and multiplying on the left-hand side by $\mathbf{T}$, we get:

$$\mathbf{K}_{2\mathrm{red}}\ddot{\mathbf{z}}(t) + \mathbf{K}_{1\mathrm{red}}\dot{\mathbf{z}}(t) + \mathbf{K}_{0\mathrm{red}}\mathbf{z}(t) = \mathbf{F}_{\mathrm{red}}(t) \tag{A10}$$

where

$$\mathbf{K}_{2\mathrm{red}} = \mathbf{T}\mathbf{K}_2\mathbf{T}^T, \quad \mathbf{K}_{1\mathrm{red}} = \mathbf{T}\mathbf{K}_1\mathbf{T}^T, \quad \mathbf{K}_{0\mathrm{red}} = \mathbf{T}\mathbf{K}_0\mathbf{T}^T, \quad \mathbf{F}_{\mathrm{red}} = \mathbf{T}\mathbf{F} \quad .$$

The reduced system can be solved with the Runge–kutta algorithm, to obtain the time evolution of z as an interpolating function of some discrete values at a number of instants. The values of the displacements vector $\mathbf{q}(t)$ can be easily obtained from Eq. (A9).

## A.2
## Element and shape functions

An important feature of the code is the automatic mesh generation. The simplest way to achieve this goal is to make use of some triangulation algorithms (like the Delauney's one), which easily allows to generate a triangular mesh for a wide class of 2-D domains. This is the main reason to consider the triangular elements as our first choice. There is a huge literature about triangular FE suitable for our requirements. The most part of the triangular elements used in structural mechanics are based upon the Reissner-Mindlin plate theory, see [2], [8–10], [12], [13] and for this reason have not been considered for the problem proposed in this paper.

Since we deal with 2-D systems governed by second- and fourth-order PDEs, the continuity requirement for the shape functions are different in each case: we need $C^0$ shape functions for second-order problems and $C^1$ shape functions for fourth-order problems. The first case presents no difficulties, since linear shape functions with three corner nodes, each one with one DOF, satisfies it. The second case shows some obstacle because in order to satisfy the $C^1$-continuity the complexity of the element (i.e. the degree of polynomial expansion and the total number of DOFs) increases enormously. This suggests the possibility of using *non-conforming* elements, i.e. elements which do not satisfy the $C^1$-continuity requirement but have proved to converge.

In fact, even if for non-conforming elements we are not sure to get a monotonic convergence, we can prove it by means of the *Patch test*. This is a means of assessing convergence for problems in which the shape functions violate continuity requirements. The test is applicable to all FEs and, if properly extended and interpreted, it can provide a *necessary* and *sufficient* condition for convergence, and also an assessment of the convergence rate. Therefore we decide to abandon conformity

The element that we found to be convenient for our work is a modification, proposed by Specht, of one of the first non-conforming elements derived by Bazeley, Cheung, Irons and Zienkiewicz, see [1]. To avoid asymmetry and gain a general formulation, let is introduce the area-coordinates, which are indeed a natural choice for triangles, Fig. (18).

We use polynomial expansion terms given in area-coordinates, e.g.

$$\alpha_1 L_1 + \alpha_2 L_2 + \alpha_3 L_3 \quad ,$$

gives the three terms of a complete linear polynomial and

$$\alpha_1 L_1 L_2 + \alpha_2 L_2 L_3 + \alpha_3 L_3 L_1 + \alpha_4 L_1^2 + \alpha_5 L_2^2 + \alpha_6 L_3^2$$

gives all six terms of quadratic one and so on.

For a nine DOFs element (three corner nodes, each one with three DOFs: the displacement $\omega$ and the slopes $\theta_x$ and $\theta_y$) any of the above terms can be used in suitable condition, remembering, however, that only nine independent functions are needed. We ensure that all the six quadratic terms are present. In addition, we select three fourth-order terms. The particular form of these is so designed that the patch-test criterion is identically satisfied. We write

$$\begin{aligned} w = {} & [L_1, L_2, L_3, L_1L_2, L_2L_3, L_3L_1, \\ & L_1^2L_2 + \frac{1}{2}L_1L_2L_3\{3(1-\mu_3)L_1 - (1+3\mu_3)L_2 + (1+3\mu_3)L_3\}, \\ & L_2^2L_3 + \frac{1}{2}L_1L_2L_3\{3(1-\mu_1)L_2 - (1+3\mu_1)L_3 + (1+3\mu_1)L_1\}, \\ & L_3^2L_1 + \frac{1}{2}L_1L_2L_3\{3(1-\mu_2)L_3 - (1+3\mu_2)L_1 + (1+3\mu_2)L_2\}]\boldsymbol{\alpha} \\ & = \mathbf{P}\boldsymbol{\alpha} \quad , \end{aligned}$$

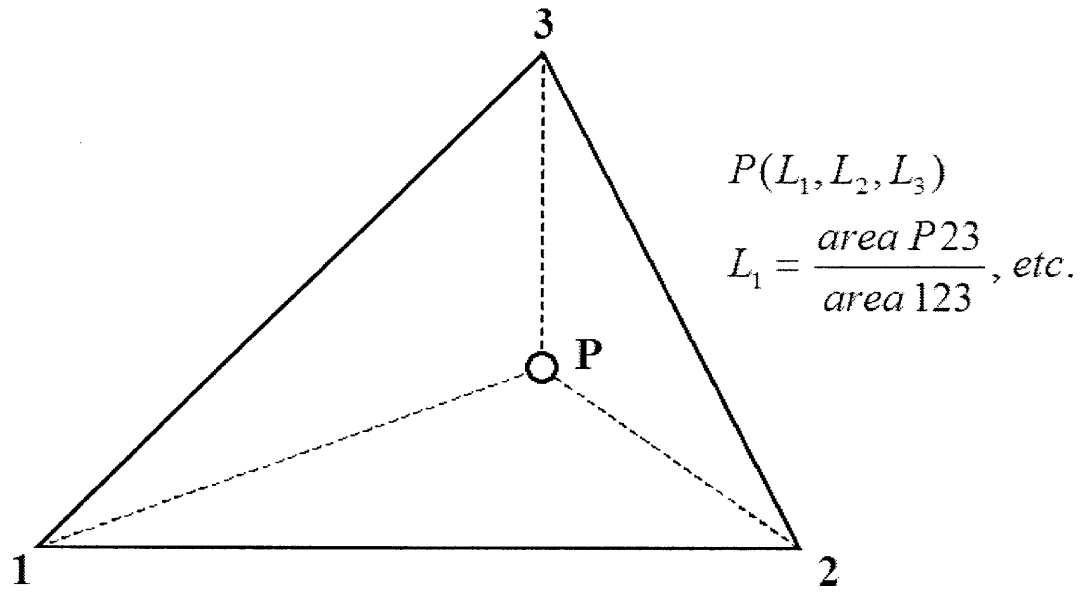

**Fig. 18.** Triangular area-coordinates

where

$$\mu_1 = \frac{l_3^2 - l_2^2}{l_1^2}, \quad \mu_2 = \frac{l_1^2 - l_3^2}{l_2^2}, \quad \mu_3 = \frac{l_2^2 - l_1^2}{l_3^2} \quad ,$$

and $l_1, l_2, l_3$ are the lenghts of the triangle sides.
On identification of nodal values and inversion, the shape functions can be written explicitly in terms of the component of the vectors **P** as

$$\mathbf{N}_i^T = \begin{bmatrix} \mathbf{P}_i - \mathbf{P}_{i+3} + \mathbf{P}_{k+3} + 2(\mathbf{P}_{i+6} - \mathbf{P}_{k+6}) \\ -b_j(\mathbf{P}_{k+6} - \mathbf{P}_{k+3}) - b_k\mathbf{P}_{i+6} \\ -c_j(\mathbf{P}_{k+6} - \mathbf{P}_{k+3}) - c_k\mathbf{P}_{i+6} \end{bmatrix},$$

where $i, j, k$ are a cyclic permutation of 1, 2, 3 and

$$b_1 = y_2 - y_3,$$

$$c_1 = x_3 - x_2,$$

where $\{x_i, y_i\}$ are the coordinates of node $i$.

The element now derived passes all the patch tests and performs excellently for all fourth-order problems, like the Krichhoff–Love plate theory.

If we deal with a certain number of coupled system, some of second-order and some of fourth-order, we can easily build the global shape functions matrix by simply alternating, in a proper way, those shape functions (for fourth-order problems) with linear shape functions (for second-order problem). Obviously, on each corner node of the triangular elements there will be defined three independent variables for each fourth-order function and one independent variable for each second-order function. calling $n_4$ and $n_2$ the number of fourth-order and second-order functions, respectively, we have that the total number of DOFs per element is

$$DOF = 3(3n_4 + n_2) \quad .$$